\documentclass[fleqn,usenatbib]{mnras}

\pdfoutput=1

\usepackage{newtxtext,newtxmath}

\usepackage[T1]{fontenc}
\usepackage{ae,aecompl}
\usepackage{tablefootnote}
\usepackage{threeparttable}

\usepackage{graphicx}	
\usepackage{amsmath}	
\usepackage{amssymb}	

\title[The ATLAS 9.0 GHz Survey of the Extended Chandra Deep Field South]{The ATLAS 9.0 GHz Survey of the Extended Chandra Deep Field South: The Faint 9.0 GHz Radio Population.}

\author[M. T. Huynh et al.]{
M. T. Huynh,$^{1,2}$\thanks{E-mail: Minh.Huynh@csiro.au}
N. Seymour,$^{3}$
R.P. Norris,$^{4,5}$
and T. Galvin$^{1}$
\\
$^{1}$CSIRO Astronomy and Space Science (CASS), PO Box 1130, Bentley WA 6102, Australia \\
$^{2}$International Centre for Radio Astronomy Research, M468, University of Western Australia, Crawley, WA 6009, Australia \\
$^{3}$International Centre for Radio Astronomy Research, Curtin University, Perth, Australia\\
$^{4}$CSIRO Astronomy and Space Science (CASS), PO Box 76, Epping, NSW 1710, Australia \\
$^{5}$Western Sydney University, Penrith Campus, Locked Bag 1797, Penrith NSW 2751
}

\date{Accepted XXX. Received YYY; in original form ZZZ}

\pubyear{2019}

\begin{document}
\label{firstpage}
\pagerange{\pageref{firstpage}--\pageref{lastpage}}
\maketitle

\begin{abstract}
We present a new image of the 9.0 GHz radio emission from the extended Chandra
Deep Field South. A total of 181 hours of integration with the Australia Telescope 
Compact Array has resulted in a 0.276 square degree image with a median sensitivity of $\sim$20 $\mu$Jy/beam rms, for a synthesised beam of 4.0 $\times$ 1.3 arcsec. We present a catalogue of the 9.0 GHz radio sources, identifying 70 source components and 55 individual radio galaxies.  Source counts derived from this sample are consistent with those reported in the literature. The observed source counts are also generally consistent with the source counts from simulations of the faint radio population. Using the wealth of multiwavelength data available for this region, we classify the faint 9 GHz population and find that 91\% are radio loud AGN, 7\% are radio quiet AGN and 2\% are star forming galaxies.  The 9.0 GHz radio sources were matched to 5.5 and 1.4 GHz sources in the literature and we find a significant fraction of flat or inverted spectrum sources, with 36\% of the 9 GHz sources having $\alpha_{5.5GHz}^{9.0GHz}$ $>$ -0.3 (for $S \propto \nu^\alpha$). This flat or inverted population is not well reproduced by current simulations of radio source populations.

\end{abstract}




\section{Introduction}

Radio continuum surveys provide important insights into galaxy evolution and cosmology \citep{norris2017}. The brightest radio sources are associated with active galactic nuclei (AGN) activity (e.g. \citealp{begelman1984}). Early radio source counts of the bright radio population provided the first evidence that objects in the universe, in this case quasi-stellar objects or QSOs,  evolve cosmologically, as the observed count differed from that expected of a steady-state or static Euclidean universe \citep{ryle1961,longair1966}.  At fainter flux density levels a flattening of slope of radio source counts was found \citep{fomalont1984,condon1984,windhorst1985}, and this has been interpreted as the appearance of a new population dominated by star-forming galaxies (e.g. \citealp{windhorst1985,condon1989,seymour2004,huynh2005}).  A growing number of studies are finding that low luminosity AGN, both radio-loud and radio-quiet, make a significant contribution to the sub-mJy population (e.g. \citealp{huynh2008,seymour2008,smolcic2008,padovani2009,padovani2011,bonzini2013}). A full understanding of the different radio AGN populations, their distribution in radio luminosity, their host galaxies properties and their cosmic evolution, is essential for understanding how galaxies evolve. 

The high frequency radio sky ($\nu \gtrsim$ 10 GHz) has not been as well studied as the radio population at lower radio wavelengths (e.g. 1.4 GHz). The small field of view of radio telescopes at high frequencies, combined with source intrinsic steep radio spectral energy distributions, means that the time required to survey an area to an equivalent depth at these high frequencies is prohibitive. The Ninth Cambridge (9C) survey at 15 GHz, covering 520 sq deg to a completeness limit of 25 mJy, was the first high-frequency survey to cover a large area of sky \citep{waldram2003}. The whole southern sky has been surveyed by the Australia Telescope 20 GHz survey (AT20, \citealp{murphy2010}), which is complete to 100 mJy. Following on from that work, a pilot survey for AT20-deep covers 5 sq deg and is almost complete to 2.5 mJy \citep{franzen2014}. The 10C survey at 15.7 GHz covered ten different fields totalling $\sim$27 sq deg complete to 1 mJy \citep{franzen2011,davies2011}, with deep areas within those fields totalling $\sim$ 12 sq deg complete to 0.5 mJy. Further 15.7 GHz observations in two 10C fields go a factor of $\sim$3 deeper, reaching $\sim$16 and $\sim$21 $\mu$Jy rms over approximately 0.6 sq deg \citep{whittam2016}. 

At the bright end ($> 5$ mJy), these surveys have found that high frequency source counts are consistent with models of the known radio populations \citep{waldram2003,massardi2008}. However more recent work, probing fainter populations, is finding that the high frequency radio population models are under-predicting the observed source counts by up to a factor of two at faint flux densities ($\lesssim 1$ mJy) \citep{davies2011,whittam2016}. This has been interpreted as a population of flat spectrum sources at faint flux densities which have not been accounted for in the models \citep{whittam2013,whittam2016}, and demonstrates that one can not extrapolate from the well-determined results at lower frequency (e.g. 1.4 GHz) to high frequency by assuming a single spectral index. This is because of the changing nature of the sources contributing to the counts at lower frequency versus that at higher frequency, as a function of flux density. Sources in the deeper 1.4 GHz imaging may not be present in the higher frequency observations and a high frequency population which is flat or inverted may be undetected at 1.4 GHz.  A flattening of the average spectral index in sub-mJy samples selected at high frequencies (>10 GHz) has been observed \citep{whittam2013,franzen2014}, and this has been observed also in faint 5 GHz selected samples \citep{prandoni2006,huynh2015}, but faint radio sources selected at 1.4 GHz or 610 MHz do not appear to exhibit a flattening in their average spectral index \citep{ibar2009}. Hence there appears to be a significant population of low-luminosity flat or inverted radio AGN detected in the deepest high frequency observations which don't appear to be in the lower frequency surveys. 

In this paper we present 9.0 GHz Australia Telescope Compact Array (ATCA) observations of the extended Chandra Deep Field South (eCDFS) by the Australia Telescope Large Area Survey (ATLAS, \citealp{norris2006}) team. 
The extensive multiwavelength data in the region, from ultra-deep Hubble Space Telescope (HST) imaging \citep{giavalisco2004,rix2004,koekemoer2011}, optical spectroscopy \citep{silverman2010,cooper2012},  to infrared Herschel imaging \citep{oliver2012}, made the eCDFS an attractive target for radio observations for studying galaxy evolution. Several epochs of 1.4 and 1.0--2.0 GHz ATCA observations have been performed by our team, reaching 14 $\mu$Jy/beam rms over 3.6 sq deg incorporating the eCDFS region \citep{franzen2015}. The 5.5 GHz observations of the central 30 $\times$ 30 arcmin eCDFS were obtained by the ATLAS team in two epochs of observations (2009-2010 and 2012) of the central $\sim 0.25$  and source counts and spectral energy distribution (SED) analysis presented in an initial set of papers (\citealp{huynh2012a,huynh2015}, hereafter refereed to as H12 and H15). Higher frequency data were taken simultaneously with the 5.5 GHz data by the ATCA, centered at 9.0 GHz. The deep and relatively wide-area of our 9.0 GHz survey presents an opportunity to study faint high-frequency radio populations.  We present our 9.0 GHz observations and the data reduction in Section 2. Section 3 details the source extraction, the 9.0 GHz catalogue and its completeness. The source counts are presented in Section 4. The nature of the faint 9.0 GHz radio population is discussed in Section 5, including the radio-loud AGN vs radio-quiet AGN nature, redshifts, radio luminosities, and source radio spectral energy distributions. 

Throughout this paper a standard $\Lambda$-CDM cosmology is adopted with parameters of $\Omega_{\rm M} = 0.29$,  $\Omega_{\rm \Lambda} = 0.71$, and a Hubble constant of 69.3 km s$^{-1}$ Mpc$^{-1}$. Spectral indices are defined as $S \propto \nu^\alpha$. 

\section{The Observations and Data Reduction}

\begin{table}
\caption{Summary of the ATCA 9 GHz observations.}  
\begin{tabular}{llcc}  
\hline
Program ID & Epoch and Date & Array & Net Integration \\ 
                   &                        &           &         Time (h)     \\ \hline
C2028      &     1,  2009 Aug 12, 14 & 6D & 13.7 \\
C2028      &     1,  2010 Jan   4 -- 15 & 6A & 91.0 \\
C2670      &     2,  2012 May 31 -- June 4 & 6A & 41.7 \\
C2670      &     3,  2012 Aug 14 -- 18 & 6D & 34.3 \\
\hline
\end{tabular}
\label{tab:obssumm}
\end{table}

\begin{figure*}
\includegraphics[width=14cm]{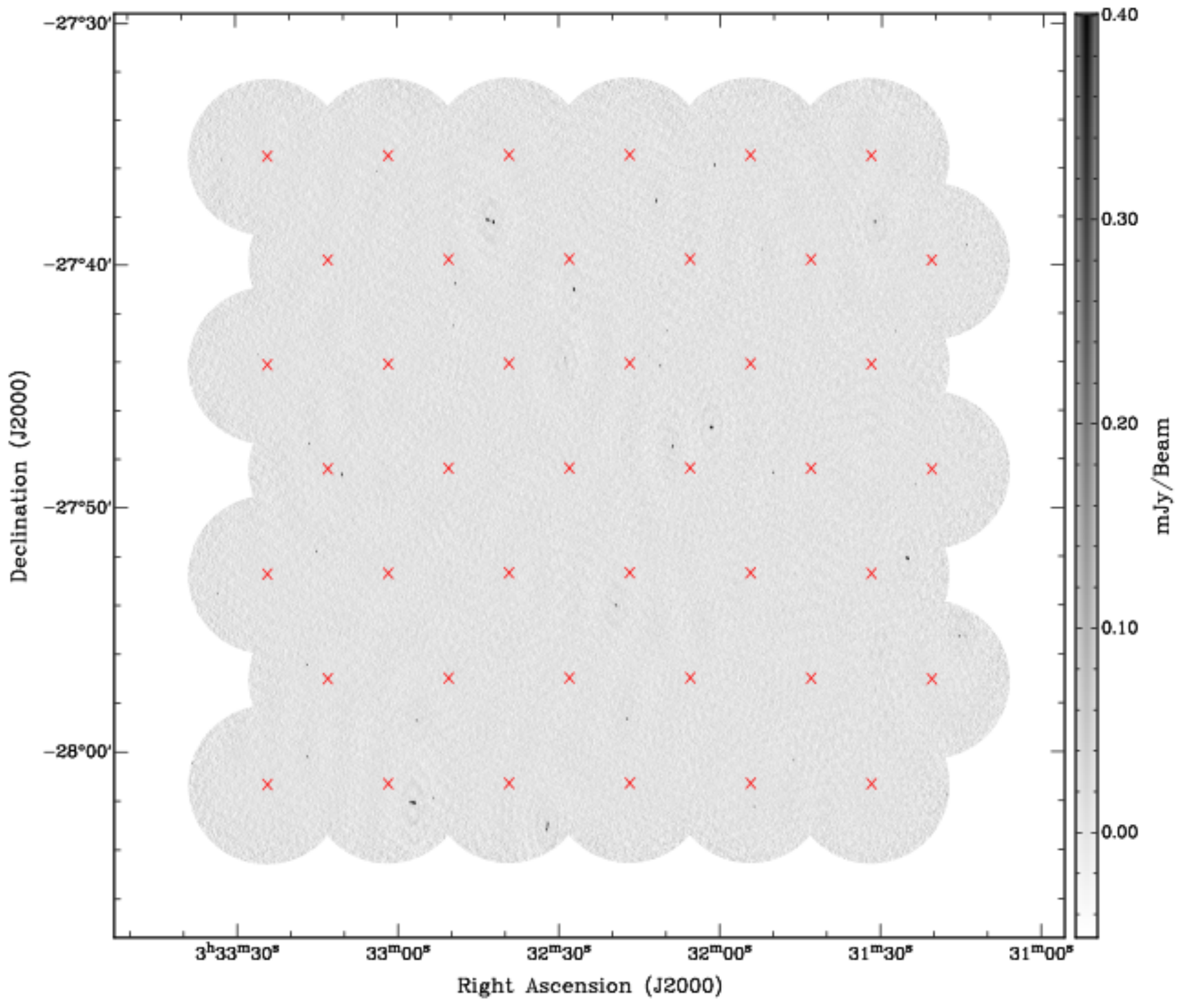}
\caption{The eCDFS 9 GHz mosaic covering 0.276 deg$^2$. The greyscale range is $-0.05$ to 0.4 mJy. The red crosses mark the 42 pointings which were combined to form the full mosaic.} 
\label{im:ecdfs_grey}
\end{figure*}

The 9 GHz observations of eCDFS were obtained with the Compact Array Broadband Backend (CABB, \citealp{wilson2011}) on the Australia Telescope Compact Array during 2009, 2010 and 2012 as part of project IDs C2028 and C2670.  CABB can observe simultaneously in two independent tunable intermediate frequency (IF) bands, each with up to 2048 MHz of bandwidth. The first IF was centred at 5.5 GHz, and those results are described in H12 and H15. The second IF was centred at 9.0 GHz.  The eCDFS region was covered with a 42 pointing hexagonal pattern with spacings set at 5 arcmin, to uniformly sample the region in IF1, 5.5 GHz.  The total integration time is summarised in Table \ref{tab:obssumm}.

The data were reduced using the Multichannel Image Reconstruction, Image Analysis and Display (MIRIAD) software package \citep{sault1995}, which is the standard reduction tool for ATCA observations. The 42 individual pointings were reduced in a similar way to the 5.5 GHz data (H15), including the use of multi-frequency synthesis. A robust weighting of 1 was used in the {\em invert} imaging step, and the individual images have a cellsize of 0.25 arcsec and size of 2800 $\times$ 2800 pixels. This extends beyond the 5\% primary beam response level of the lower end of the 9.0 GHz band, to capture the full field of view. 

Two iterations of self calibration was employed. The first iteration of self calibration used a model determined from 50 mfclean iterations, and the second with the model set by cleaning to 4$\sigma$. The final images after the second self calibration step were cleaned to 4$\sigma$. The images from the individual pointings were restored with the same beam, the average beam of the 42 pointings,  4.0 $\times$ 1.3 arcsec, and then mosaiced together using the task {\em linmos}, which applies the primary beam correction. The wideband primary beam correction is performed using the spectral index plane which is produced in the multi-frequency synthesis. 

The final mosaic is shown in Figure \ref{im:ecdfs_grey}. Outer regions further than $\sim$ 3.3 arcmin from the edge pointings, i.e. have $\gtrsim$ 0.5 FWHM primary beam response, were removed to minimise primary beam effects at the edge of the image. The image edge would otherwise have high levels of non-Gaussian noise which would increase the rate of detection of spurious sources by source finders. This final mosaic has an area of 0.276 square degrees. 

\subsection{Image Analysis: Sensitivity and Clean Bias}

To examine the noise properties of the final mosaic we ran BANE \citep{hancock2018} to produce a rms map. BANE produces an rms map by calculating statistics over a square region of the image, in this case 55 $\times$ 55 arcsec or 23 times the geometric mean of the restoring beam, and 14 times the beam in the Dec direction. BANE performs 3 iterations of sigma clipping to avoid contamination from source pixels in the image. The box size chosen for rms calculations must be large enough to avoid bias from real sources yet small enough to reflect local variations in the rms. The box size in this work is consistent with the size recommended by \citep{huynh2012b}, 10 to 20 times the beamsize. 

\begin{figure}
\includegraphics[width=\columnwidth]{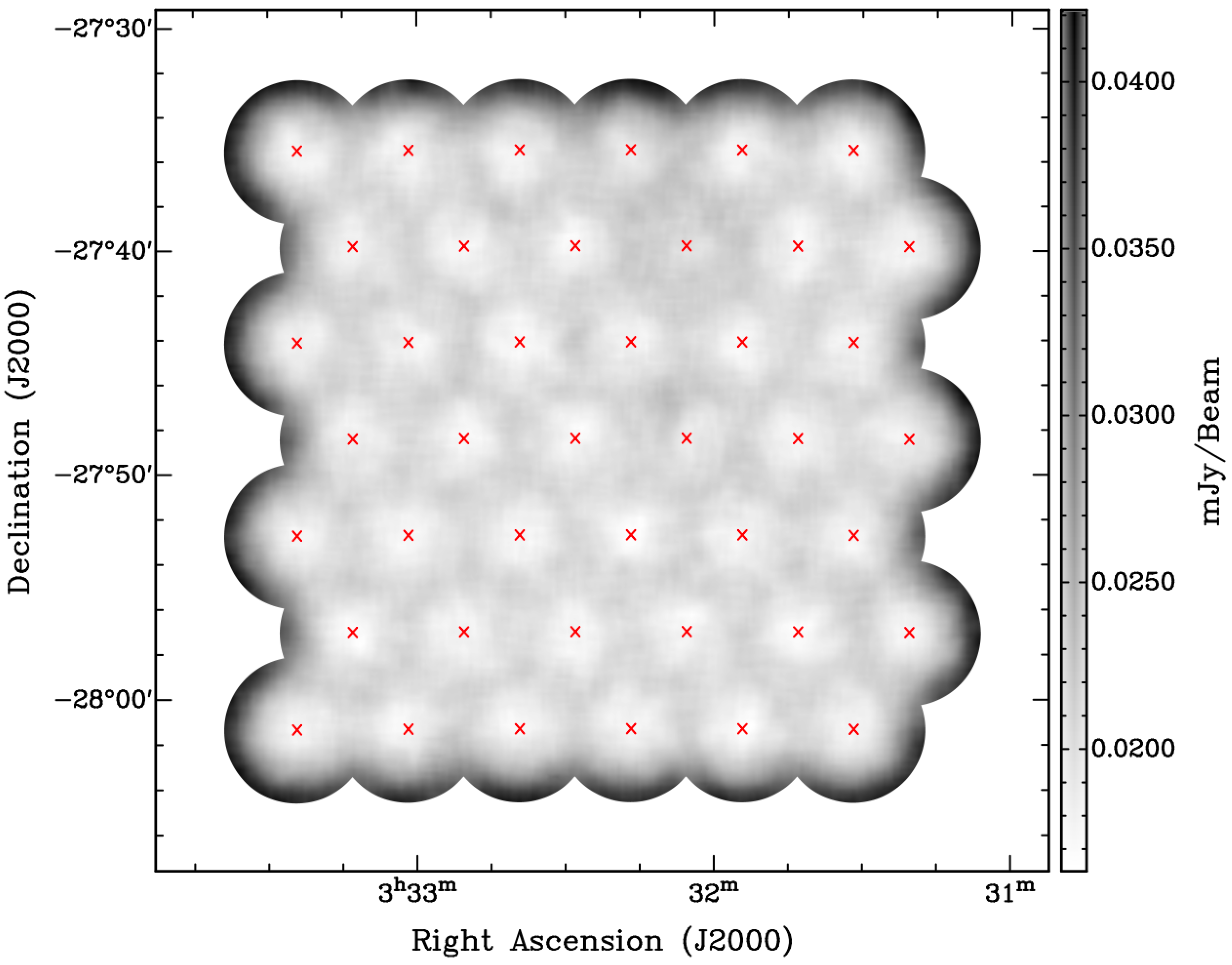}
\caption{The noise image produced by BANE. The red crosses mark the 42 pointings of the mosaic. In the interior of the mosaic the noise ranges from $\sim$17 $\mu$Jy at the pointing centres to  $\sim$21 $\mu$Jy in areas furthest from the pointing centres. } 
\label{im:noiseim_grey}
\end{figure}

\begin{figure}
\includegraphics[width=\columnwidth]{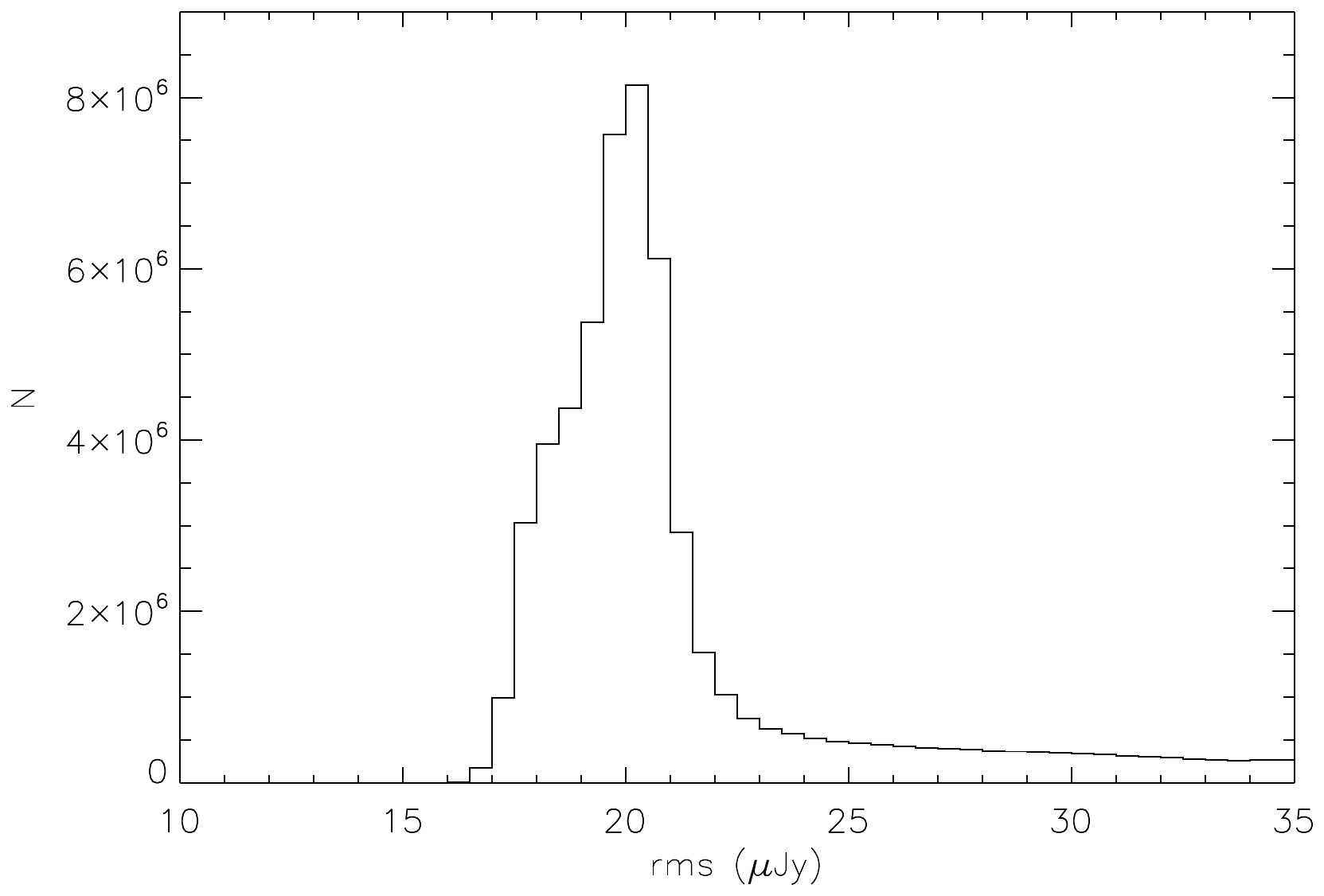}
\caption{The distribution of the pixels of the noise image produced by BANE. The peak is at 20 $\mu$Jy. There is a tail of pixels at higher noise levels ($>$25 $\mu$Jy rms) due to noisy regions around bright sources and the edge of the mosaic which has a lower primary beam response.} 
\label{im:noise_hist}
\end{figure}

The spacing of the pointings was optimised for 5.5 GHz, not 9.0 GHz, however we find the noise varies in the interior of the final mosaic by only $\sim$20\%. At the centre of the pointings the noise is about 17 to 19 $\mu$Jy, and at the areas of least overlap, the noise increases to roughly 20 to 21 $\mu$Jy (see Figure \ref{im:noiseim_grey}).  The noise at the edge of the mosaic, where the primary beam response level is 0.5 and there are no overlapping pointings, is $\sim$40 $\mu$Jy. The distribution of the pixels in the rms image is shown in Figure \ref{im:noise_hist}. The distribution has a peak of about 20 $\mu$Jy, consistent with the median of the noise image, which is 20.18 $\mu$Jy. 

Flux can be redistributed on to noise peaks during the cleaning process. This so-called clean bias is generally only a problem for snapshot observations where uv coverage is poor, or when cleaning deeply to a low threshold level (e.g. $< 3\sigma$). Similar to our previous work in H15 we performed clean bias simulations to quantify this issue. Point sources were injected into the {\em uv} visibility data of three representative pointings at random positions. The chosen pointings were the pointings of lowest and highest time integration, and one with close to the average time integration. The {\em uv} visibility data was then imaged using the same cleaning depth as the final images. The injected flux densities were compared to output ones to determine the clean bias. The simulated sources were injected one at a time to avoid overlapping with real sources in the data, and the process repeated 5000 times to obtain good statistics.  The median clean bias is negligible for brighter sources ($>$0.5 mJy) and only 6 to 7\% for the faintest sources ($\sim$100 $\mu$Jy). Clean bias is therefore not a significant issue.

\section{Source Extraction}

 Following our previous work in H12 and H15, we use the MIRIAD task {\em sfind} to detect radio sources in the image. {\em sfind} implements a false detection rate algorithm \citep{miller2001}. This compares the distribution of image pixels to that of a noise-only image to return a list of sources above a false detection threshold. In common source finders (e.g. Aegean, \citealp{hancock2012,hancock2018}) the user set threshold is a S/N limit, but with {\em sfind} the user sets the fraction of sources which are allowed to be false.

As in H12 and H15, {\em sfind} was run with `rmsbox' set to 10 times the synthesised beamwidth (90 pixels) and `alpha' set to 1. For pure Gaussian noise setting `alpha' to 1 would return a list of sources which is 99\% reliable, however the noise is correlated between pixels for the radio image and there is primary beam variation in the noise. Also, for pure Gaussian noise, this would be equivalent to a threshold setting of about 4.9$\sigma$. We find this detection method reliable - performing {\em sfind} on the negative image returns no detected sources.

\begin{figure*}
\includegraphics[width=4.9cm]{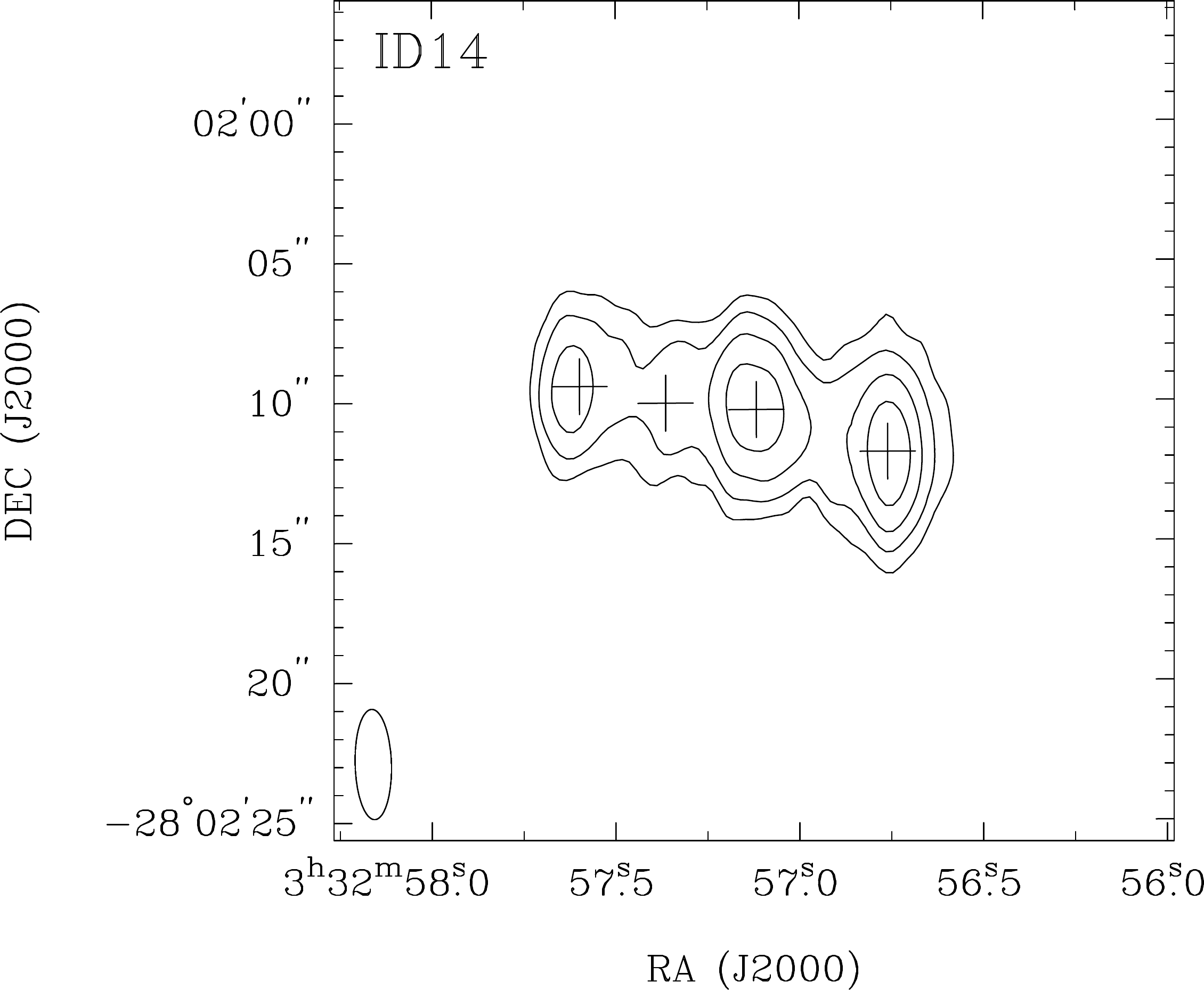}
\includegraphics[width=4.9cm]{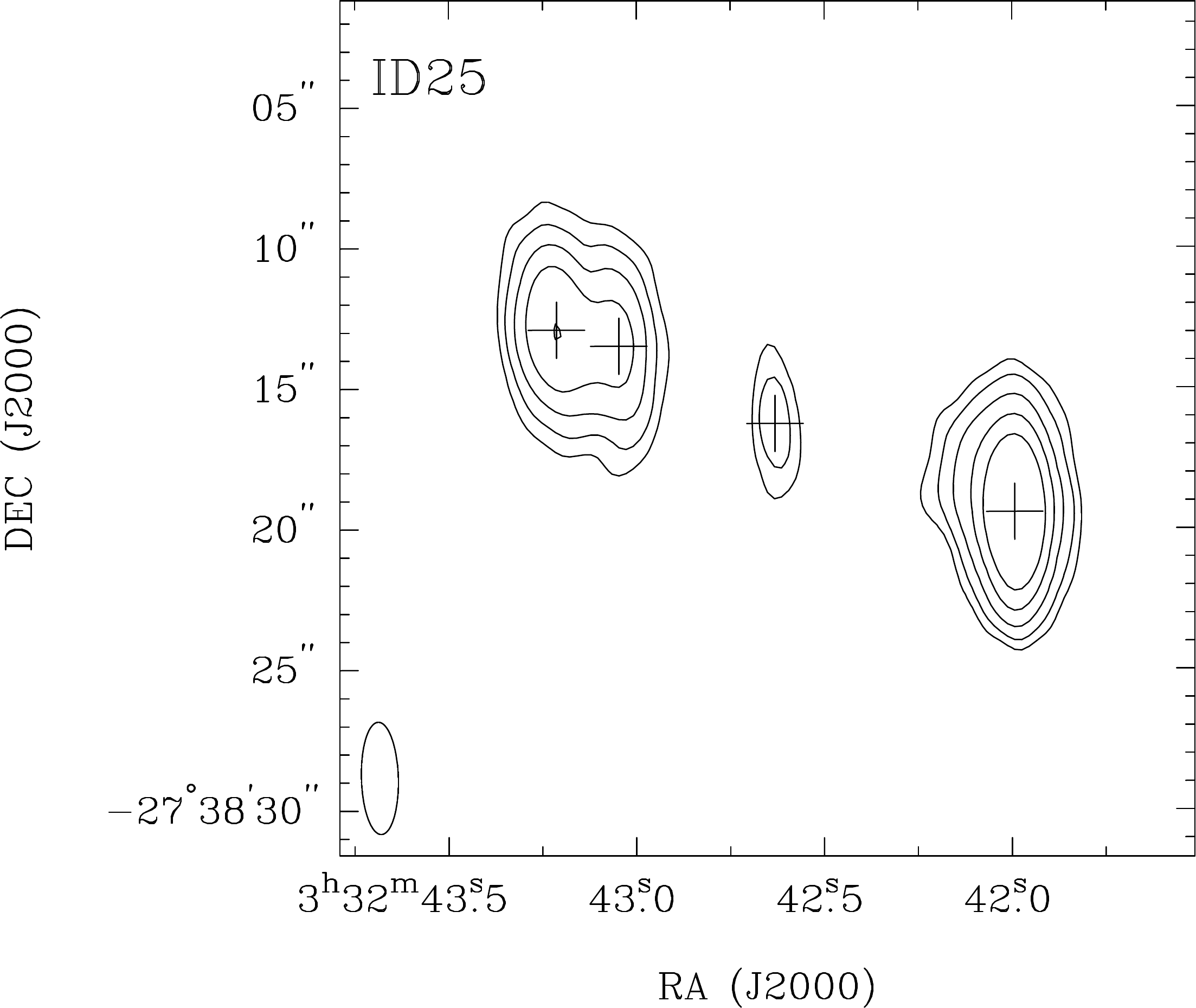}
\includegraphics[width=4.9cm]{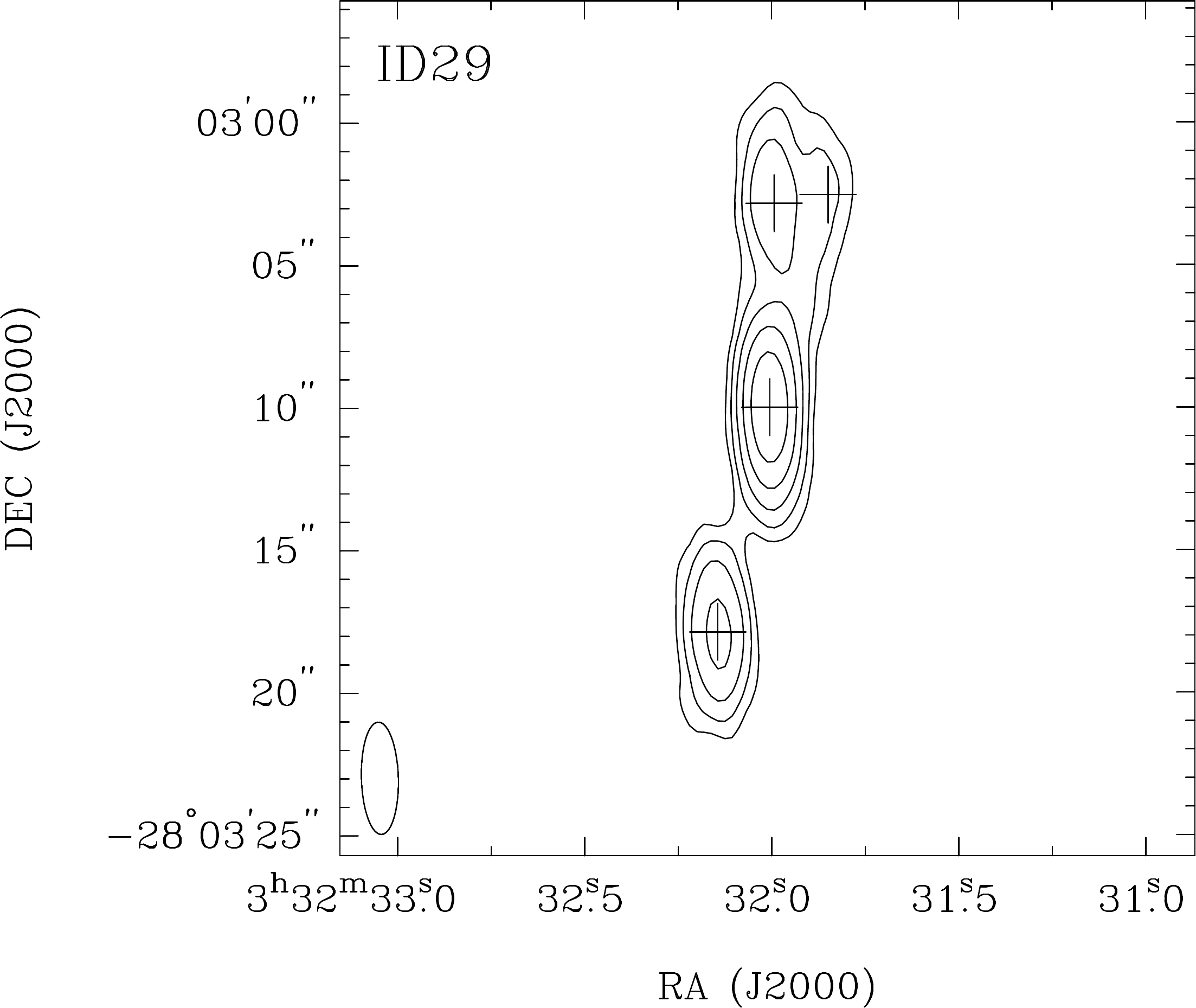}
\includegraphics[width=4.9cm]{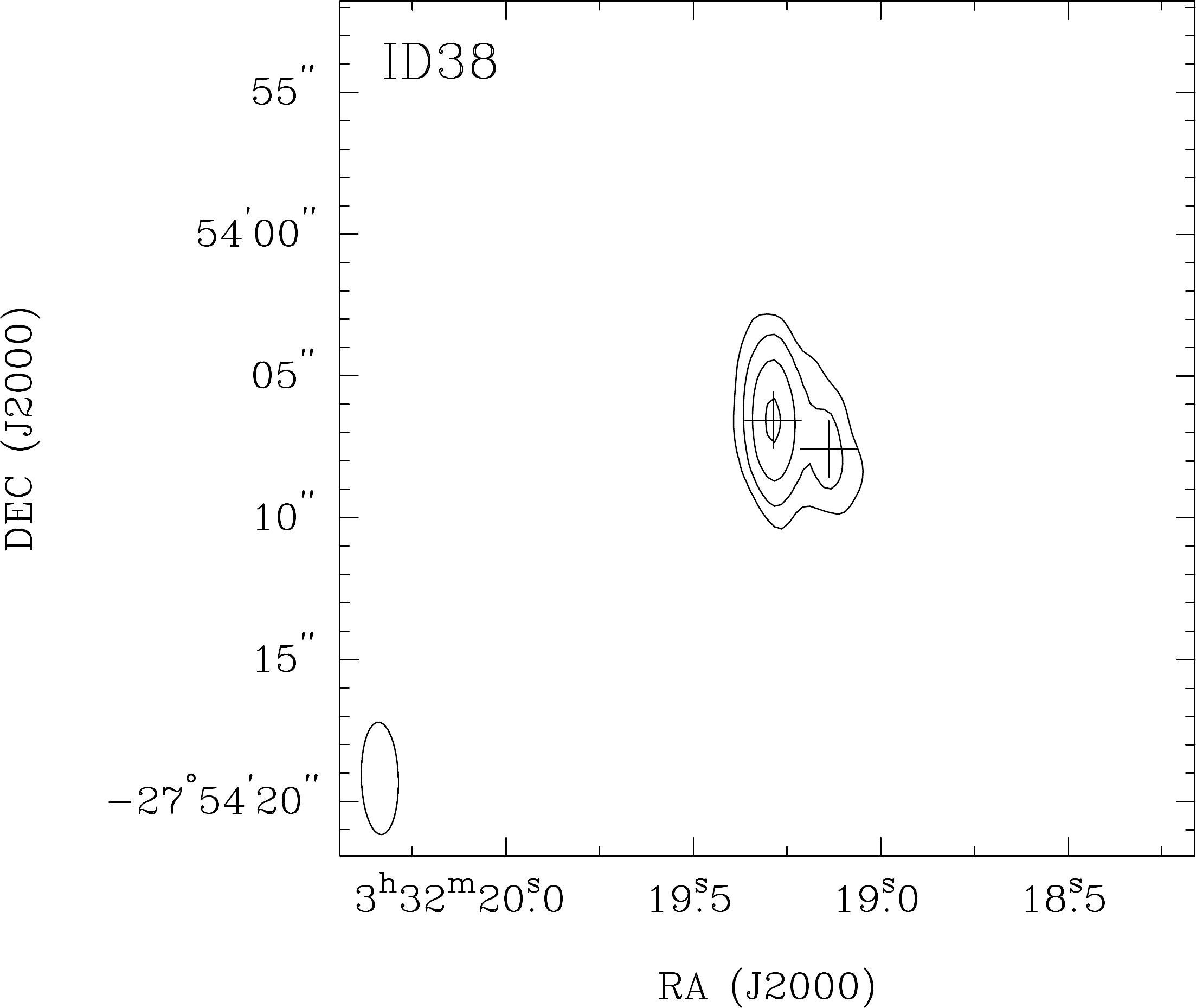}
\includegraphics[width=4.9cm]{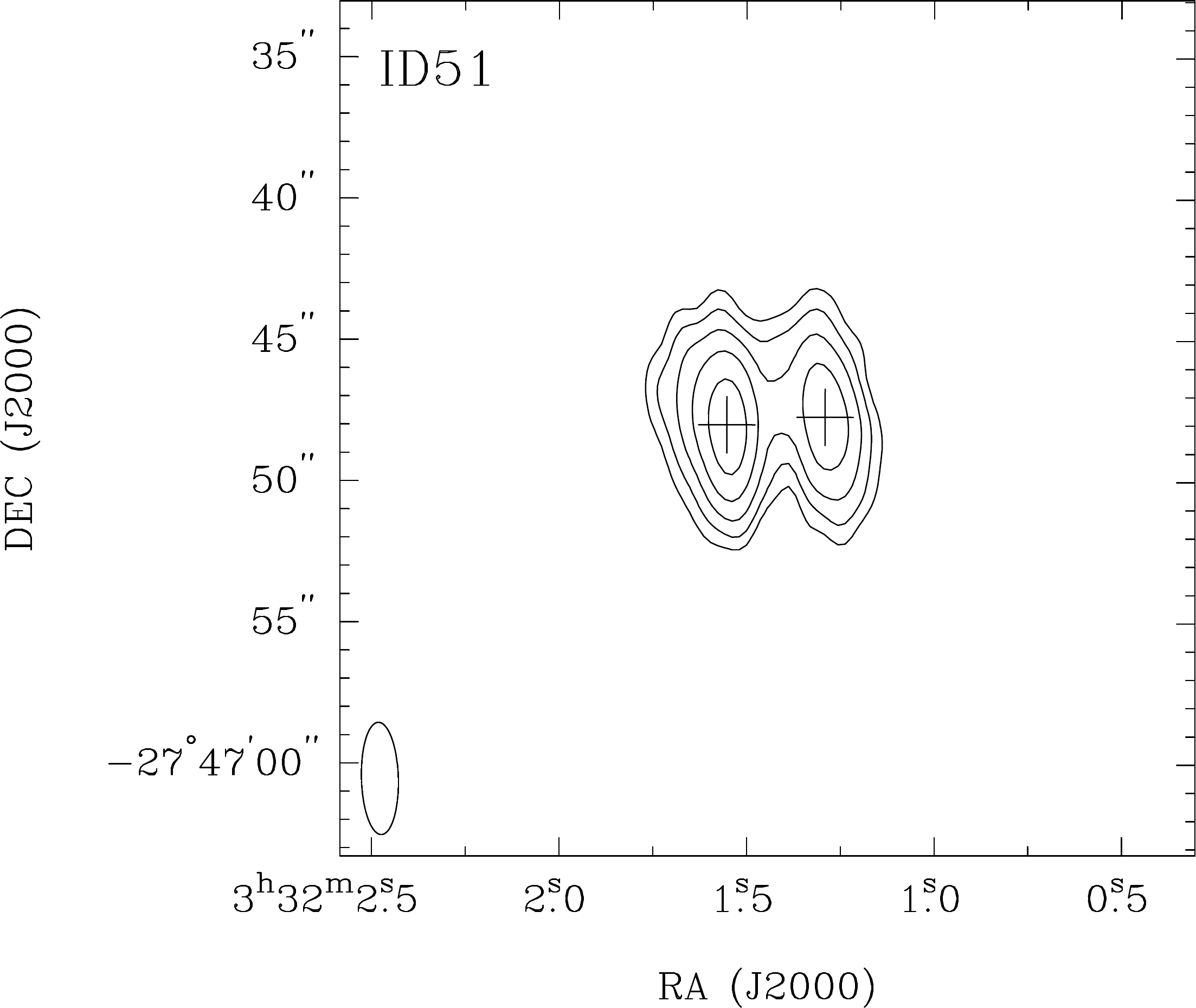}
\includegraphics[width=4.9cm]{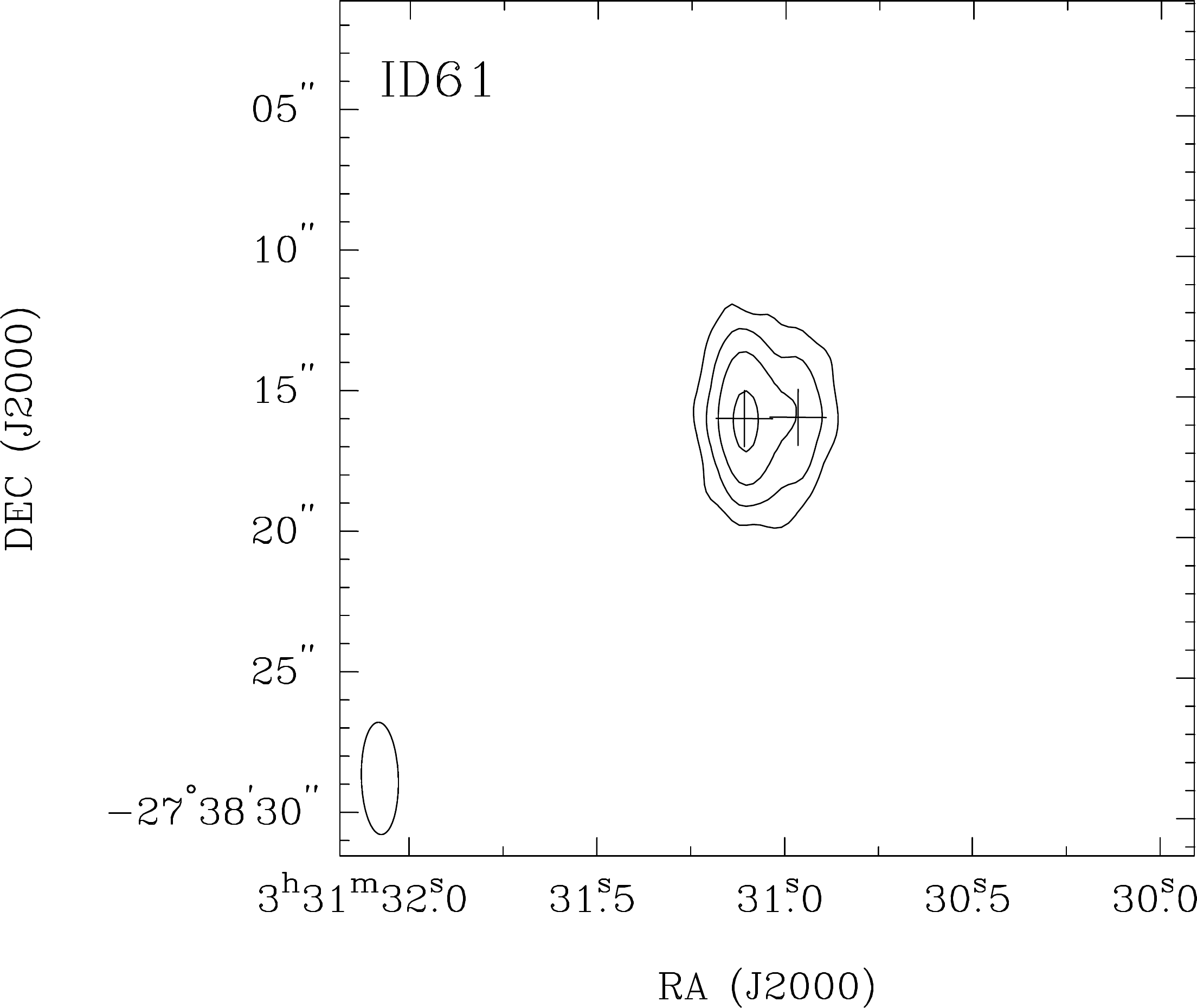}
\includegraphics[width=4.9cm]{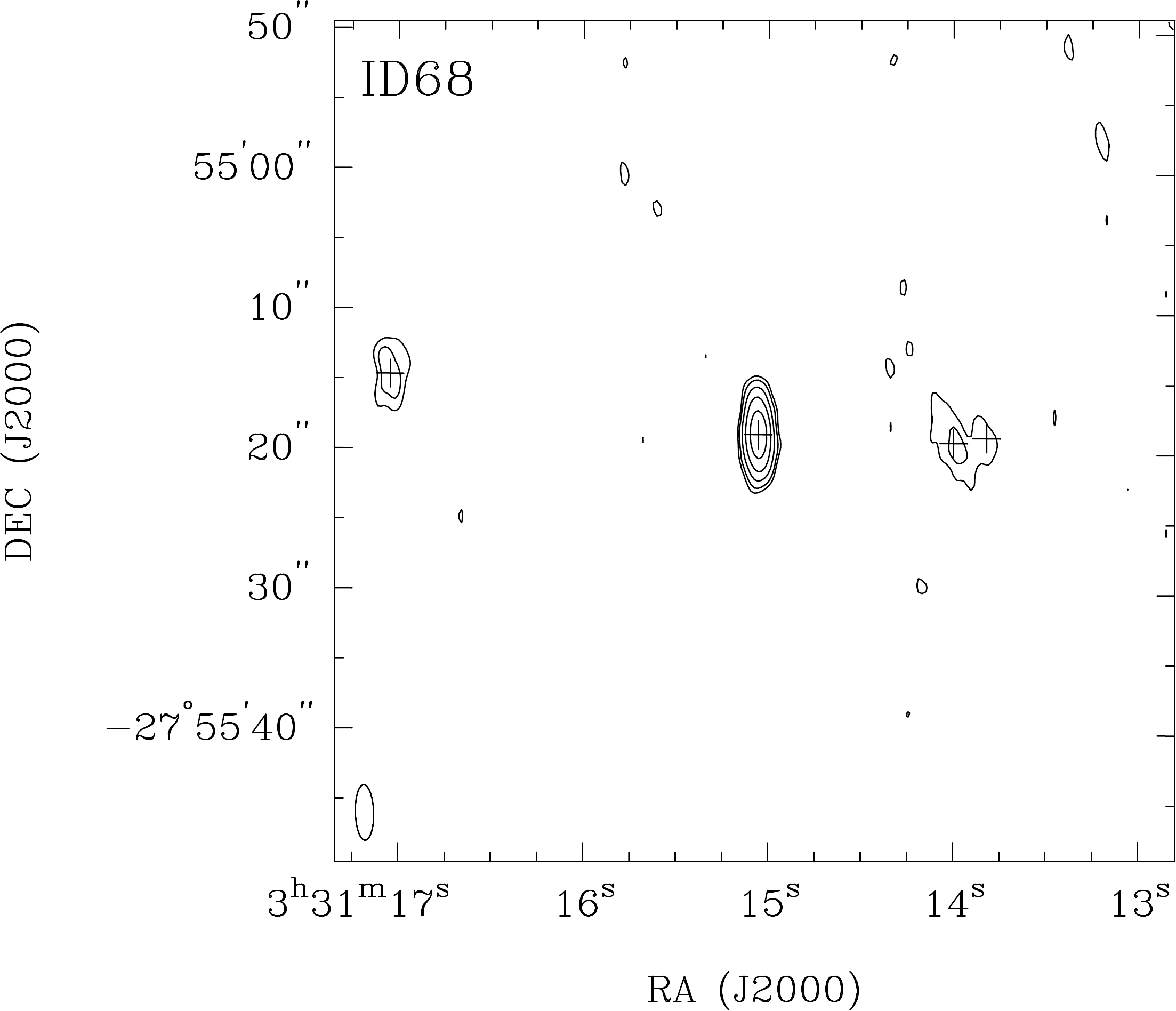}
\caption{Contour images of the multiple-component radio sources in the catalogue. The images are 30 $\times$ 30 arcsec in size, except for ID 68, which is 1 $\times$ 1 arcmin.  The contour levels are set at 5, 10, 20, 40 and 80 times the local noise level. However ID 68 also have a 3 sigma contour to show the faint lobes. The synthesized beam is shown in the bottom left corner. Crosses mark the positions of the catalogued components.}
 \label{fig:multicomp}
\end{figure*}

The {\em sfind} task returned 72 detected radio source components. Two of these are at the edge of the image with only half the source inside the final mosaic. One of these border sources is a clear bright radio source, but the other is not detected in more sensitive 1.4 GHz VLA \cite{miller2013} and 5.5 GHz ATCA data (H15). We removed the first as it is not fully within our mosaic and we removed the second as it is likely spurious. One spurious detection out of 72 is consistent with the false detection threshold we set (1\%), and still very reliable. Each of the remaining 70 {\em sfind} source components was then individually fit as both a point source and a Gaussian using MIRIAD task {\em imfit}.  Seven complex multiple-component sources were identified via visual inspection (see Figure \ref{fig:multicomp}), and these cases were confirmed with inspection of deeper 1.4 GHz VLA data \citep{miller2013} and 5.5 GHz ATCA data (H15). These sources have core-lobe or lobe-lobe radio AGN morphology and were confirmed as components of a single radio source or radio galaxy using optical/NIR of the field. Where necessary these sources were re-fitted as multiple Gaussians with {\em imfit} and the components are listed individually in the final catalogue. There are 55 individual radio sources, or radio galaxies, and 70 source components in the final catalogue. 

\subsection{Deconvolution}
\label{sec:deconv}

Following H12 and H15, we use the ratio of integrated flux density to the fitted peak Gaussian flux (see Equation 1 of H12) to determine if a source is resolved. Whether a source is successfully deconvolved depends on the S/N ratio of the source as well as the synthesised beam-size (or resolution of the image). 

\begin{figure}
\includegraphics[width=8cm]{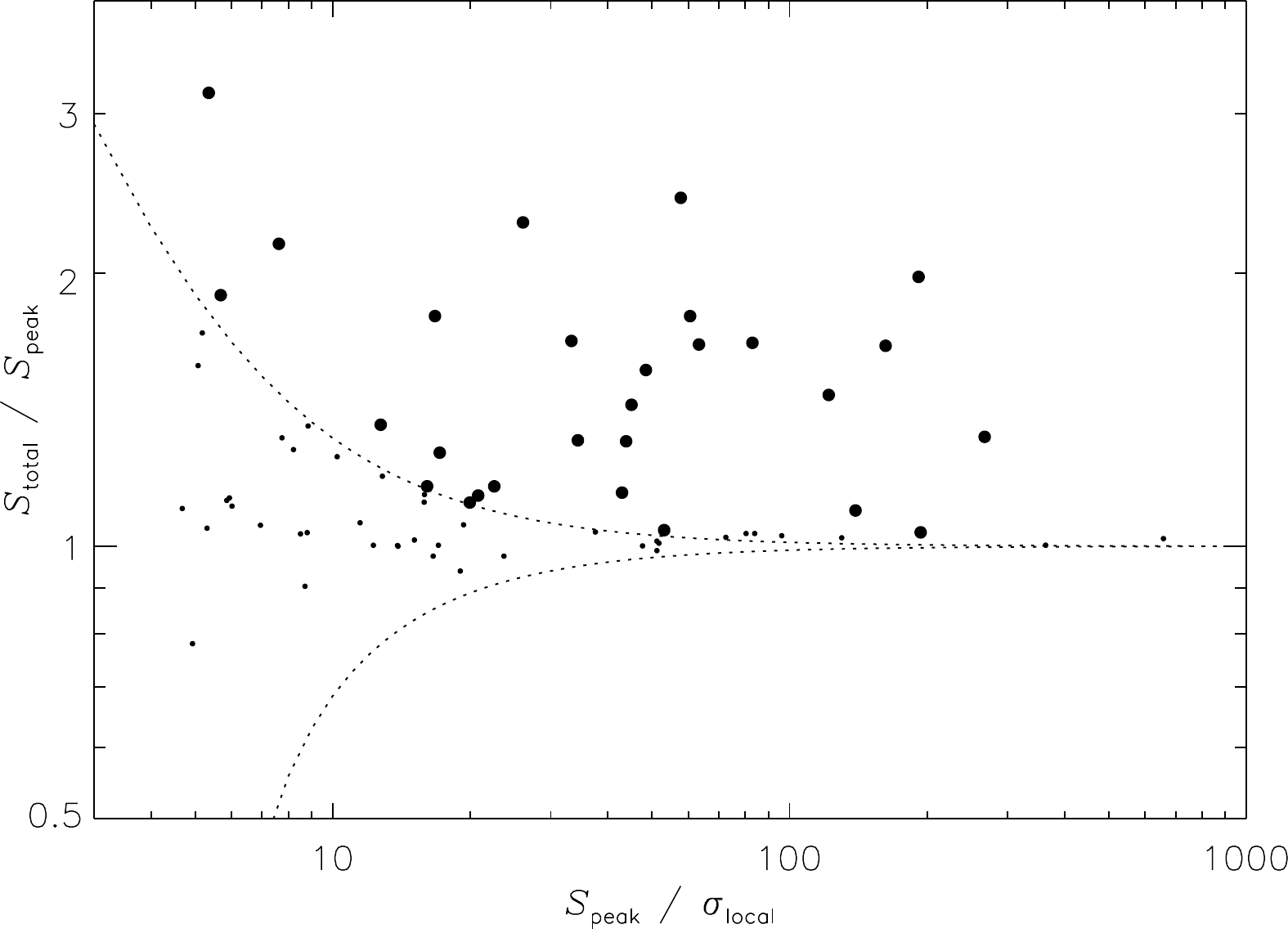}
\caption{The integrated flux density ($S_{\rm total}$) to peak flux density ($S_{\rm peak}$) ratio as a function of source signal to noise ($S_{\rm peak}/\sigma$). The dotted lines shows the upper and lower envelopes of the flux ratio distribution that contain the unresolved sources. The large dots indicate sources which are deconvolved successfully and considered resolved.} 
\label{im:resolved}
\end{figure}

We show the total integrated flux density to peak flux density ratio from the {\em imfit} Gaussian fits as a function of the S/N in Figure \ref{im:resolved}. Assuming that sources with $ S_{\rm int} / S_{\rm peak} < 1$ are due to fitting uncertainties from noise, we can define an envelope above which sources can be considered resolved:
\begin{equation}
   S_{\rm int} / S_{\rm peak} = 1 + a / (S_{\rm peak} / \sigma)^b \;.
   \end{equation}
As in H12 and H15, we take $a = 10$ and $b = 1.5$ and this is shown as the upper line in (Figure \ref{im:resolved}). The lower line is just Equation 1 simply mirrored across $ S_{\rm int} / S_{\rm peak} = 1$. It encompasses all the sources with $ S_{\rm int} / S_{\rm peak} < 1$, so we find this line is a good representation of the deconvolution limit. We also require that $  S_{\rm int} / S_{\rm peak}  > 1.035$ so that bright sources must have a significant extension, greater than the fitting uncertainty, to be deconvolved. We find 28/70 (40\%) components are resolved.

\subsection{The Source Catalogue}

The source catalogue is presented in Table \ref{tab:catalog}. Point source measurements are given for unresolved sources. For the remaining sources we provide the peak and integrated source flux density and deconvolved source sizes from the Gaussian fits. Internal fitting errors shown in Table \ref{tab:catalog} dominate for the majority of sources, which are low S/N, while absolute calibration errors dominate for high S/N sources.

Column (1) - running source ID. A letter, such as `a', `b', etc., indicates a component of a multi-component source.

Column (2) - Source IAU name

Columns (3) and (4) - Source position: Right Ascension and Declination (J2000)

Column (5) - Source peak flux density ($\mu$Jy)

Column (6) - Uncertainty in source peak flux density ($\mu$Jy)

Column (7) - Integrated flux density ($\mu$Jy). Zero indicates sources is not resolved and hence no integrated flux density is given.

Column (8) - Resolved flag. Zero indicates source is not resolved.

Column (9) - Deconvolved major axis (arcsec). Zero indicates source is not resolved. 

Column (10) - Deconvolved minor axis (arcsec). Zero indicates source is not resolved.

Column (11) - Deconvolved position angle (degrees), measured from north through east. Zero indicates source is not resolved.

Column(12) - Local noise level, rms, in $\mu$Jy.

\begin{table*}\caption{The ATLAS 9.0 GHz Catalogue}  
{\scriptsize\begin{tabular}{llllrrrrrrrr}  
\hline
ID & IAU name & RA & Dec  & $S_{p}$ & d$S_{p}$ & $S_{int}$ & Resolved & Decon & Decon & Decon & $\sigma_{local}$ \\  
     &            &   (J2000) &  (J2000)   & ($\mu$Jy)  & ($\mu$Jy)  & ($\mu$Jy) & Flag & Bmajor & Bminor & PA & ($\mu$Jy) \\ \hline

      1 &   ATCDFS9 J0033338.36-280029.6 & 03:33:38.3 & -28:00:29.6 &        375 &        40 &          0 & 0 &   0.00 &   0.00 &    0.0 &       43.7 \\
      2 &   ATCDFS9 J0033333.42-275332.4 & 03:33:33.4 & -27:53:32.4 &        386 &        33 &          0 & 0 &   0.00 &   0.00 &    0.0 &       27.8 \\
      3 &   ATCDFS9 J0033332.56-273539.0 & 03:33:32.5 & -27:35:39.0 &        324 &        25 &        441 & 1 &   3.12 &   0.48 &   -2.1 &       25.4 \\
      5 &   ATCDFS9 J0033325.86-274343.1 & 03:33:25.8 & -27:43:43.1 &        277 &        23 &          0 & 0 &   0.00 &   0.00 &    0.0 &       18.2 \\
      6 &   ATCDFS9 J0033316.73-275630.4 & 03:33:16.7 & -27:56:30.4 &        859 &        23 &          0 & 0 &   0.00 &   0.00 &    0.0 &       18.0 \\
      7 &   ATCDFS9 J0033316.75-280016.0 & 03:33:16.7 & -28:00:16.0 &       1034 &        17 &          0 & 0 &   0.00 &   0.00 &    0.0 &       20.3 \\
      8 &   ATCDFS9 J0033316.35-274724.9 & 03:33:16.3 & -27:47:24.9 &        981 &        22 &          0 & 0 &   0.00 &   0.00 &    0.0 &       18.9 \\
      9 &   ATCDFS9 J0033314.99-275151.4 & 03:33:14.9 & -27:51:51.4 &        805 &        23 &          0 & 0 &   0.00 &   0.00 &    0.0 &       21.0 \\
     10 &   ATCDFS9 J0033310.19-274842.2 & 03:33:10.1 & -27:48:42.2 &       6749 &        27 &          0 & 0 &   0.00 &   0.00 &    0.0 &       18.5 \\
     11 &   ATCDFS9 J0033309.73-274801.6 & 03:33:09.7 & -27:48:01.6 &         95 &        20 &          0 & 0 &   0.00 &   0.00 &    0.0 &       18.1 \\
     12 &   ATCDFS9 J0033308.17-275033.6 & 03:33:08.1 & -27:50:33.6 &        235 &        25 &          0 & 0 &   0.00 &   0.00 &    0.0 &       19.9 \\
     13 &   ATCDFS9 J0033303.73-273611.4 & 03:33:03.7 & -27:36:11.4 &        427 &        26 &          0 & 0 &   0.00 &   0.00 &    0.0 &       18.3 \\
    14A &   ATCDFS9 J0033257.59-280209.4 & 03:32:57.5 & -28:02:09.4 &        540 &        60 &       1230 & 1 &   2.47 &   1.72 &  -65.0 &       20.7 \\
    14B &   ATCDFS9 J0033257.36-280210.0 & 03:32:57.3 & -28:02:10.0 &        356 &        73 &          0 & 0 &   0.00 &   0.00 &    0.0 &       20.9 \\
    14C &   ATCDFS9 J0033257.11-280210.2 & 03:32:57.1 & -28:02:10.2 &       1213 &        58 &       2941 & 1 &   2.81 &   1.22 &   75.0 &       21.0 \\
    14D &   ATCDFS9 J0033256.76-280211.8 & 03:32:56.7 & -28:02:11.8 &       1290 &        67 &       2315 & 1 &   1.96 &   1.65 &    2.1 &       21.3 \\
     19 &   ATCDFS9 J0033256.47-275848.4 & 03:32:56.4 & -27:58:48.4 &       1042 &        20 &       1086 & 1 &   0.41 &   0.08 &   69.6 &       19.6 \\
     20 &   ATCDFS9 J0033253.34-280159.5 & 03:32:53.3 & -28:01:59.5 &        467 &        22 &        531 & 1 &   1.80 &   0.31 &    8.2 &       22.4 \\
     21 &   ATCDFS9 J0033252.07-274425.7 & 03:32:52.0 & -27:44:25.7 &        212 &        32 &          0 & 0 &   0.00 &   0.00 &    0.0 &       21.0 \\
     22 &   ATCDFS9 J0033249.43-274235.3 & 03:32:49.4 & -27:42:35.3 &        430 &        22 &        481 & 1 &   1.36 &   0.12 &  -19.4 &       21.6 \\
     23 &   ATCDFS9 J0033249.19-274050.7 & 03:32:49.1 & -27:40:50.7 &       1572 &        27 &          0 & 0 &   0.00 &   0.00 &    0.0 &       18.5 \\
    25A &   ATCDFS9 J0033243.21-273812.9 & 03:32:43.2 & -27:38:12.9 &       1851 &        40 &       3104 & 1 &   1.89 &   1.10 &   50.9 &       22.3 \\
    25B &   ATCDFS9 J0033243.04-273813.5 & 03:32:43.0 & -27:38:13.5 &       1082 &        47 &       1694 & 1 &   2.33 &   1.17 &    8.6 &       22.3 \\
    25C &   ATCDFS9 J0033242.63-273816.2 & 03:32:42.6 & -27:38:16.2 &        371 &        22 &          0 & 0 &   0.00 &   0.00 &    0.0 &       22.4 \\
    25D &   ATCDFS9 J0033241.99-273819.4 & 03:32:41.9 & -27:38:19.4 &       5976 &        22 &       7892 & 1 &   1.36 &   0.97 &    8.9 &       22.4 \\
    29A &   ATCDFS9 J0033232.14-280317.8 & 03:32:32.1 & -28:03:17.8 &       1228 &        28 &       1604 & 1 &   1.69 &   0.87 &    4.7 &       28.0 \\
    29B &   ATCDFS9 J0033232.00-280309.9 & 03:32:32.0 & -28:03:09.9 &       3734 &        27 &       4090 & 1 &   1.31 &   0.32 &   -8.2 &       26.8 \\
    29C &   ATCDFS9 J0033231.99-280302.8 & 03:32:31.9 & -28:03:02.8 &        855 &        68 &       1440 & 1 &   3.24 &   1.07 &    7.3 &       25.7 \\
    29D &   ATCDFS9 J0033231.84-280302.5 & 03:32:31.8 & -28:03:02.5 &        315 &        70 &          0 & 0 &   0.00 &   0.00 &    0.0 &       25.7 \\
     33 &   ATCDFS9 J0033231.55-275029.1 & 03:32:31.5 & -27:50:29.1 &        169 &        14 &          0 & 0 &   0.00 &   0.00 &    0.0 &       20.3 \\
     34 &   ATCDFS9 J0033228.82-274355.9 & 03:32:28.8 & -27:43:55.9 &        366 &        24 &          0 & 0 &   0.00 &   0.00 &    0.0 &       21.7 \\
     35 &   ATCDFS9 J0033226.97-274106.9 & 03:32:26.9 & -27:41:06.9 &       3166 &        20 &       5269 & 1 &   3.47 &   0.77 &   16.6 &       19.5 \\
     36 &   ATCDFS9 J0033224.31-280114.6 & 03:32:24.3 & -28:01:14.6 &        119 &        21 &          0 & 0 &   0.00 &   0.00 &    0.0 &       19.2 \\
     37 &   ATCDFS9 J0033221.72-280154.1 & 03:32:21.7 & -28:01:54.1 &        118 &        12 &          0 & 0 &   0.00 &   0.00 &    0.0 &       19.3 \\
    38A &   ATCDFS9 J0033219.28-275406.5 & 03:32:19.2 & -27:54:06.5 &        853 &        35 &        978 & 1 &   1.07 &   0.62 &    2.8 &       19.9 \\
    38B &   ATCDFS9 J0033219.13-275407.5 & 03:32:19.1 & -27:54:07.5 &        277 &        28 &          0 & 0 &   0.00 &   0.00 &    0.0 &       19.9 \\
     40 &   ATCDFS9 J0033218.02-274718.3 & 03:32:18.0 & -27:47:18.3 &        288 &        21 &          0 & 0 &   0.00 &   0.00 &    0.0 &       20.8 \\
     41 &   ATCDFS9 J0033217.05-275846.5 & 03:32:17.0 & -27:58:46.5 &       1628 &        36 &          0 & 0 &   0.00 &   0.00 &    0.0 &       19.9 \\
     42 &   ATCDFS9 J0033215.97-273438.1 & 03:32:15.9 & -27:34:38.1 &        181 &        22 &          0 & 0 &   0.00 &   0.00 &    0.0 &       21.3 \\
     43 &   ATCDFS9 J0033213.90-275000.6 & 03:32:13.9 & -27:50:00.6 &        144 &        29 &          0 & 0 &   0.00 &   0.00 &    0.0 &       20.8 \\
     44 &   ATCDFS9 J0033213.49-274953.6 & 03:32:13.4 & -27:49:53.6 &         89 &        24 &          0 & 0 &   0.00 &   0.00 &    0.0 &       20.7 \\
     45 &   ATCDFS9 J0033213.04-274351.5 & 03:32:13.0 & -27:43:51.5 &        101 &        18 &        191 & 1 &   2.45 &   0.79 &   52.5 &       17.8 \\
     46 &   ATCDFS9 J0033211.65-273726.3 & 03:32:11.6 & -27:37:26.3 &      14823 &       101 &          0 & 0 &   0.00 &   0.00 &    0.0 &       22.3 \\
     47 &   ATCDFS9 J0033211.01-274053.7 & 03:32:11.0 & -27:40:53.7 &        185 &        23 &          0 & 0 &   0.00 &   0.00 &    0.0 &       21.2 \\
     48 &   ATCDFS9 J0033210.92-274415.2 & 03:32:10.9 & -27:44:15.2 &       1860 &        27 &          0 & 0 &   0.00 &   0.00 &    0.0 &       19.1 \\
     49 &   ATCDFS9 J0033209.71-274248.6 & 03:32:09.7 & -27:42:48.6 &        382 &        29 &          0 & 0 &   0.00 &   0.00 &    0.0 &       20.8 \\
     50 &   ATCDFS9 J0033208.67-274734.6 & 03:32:08.6 & -27:47:34.6 &       4247 &        22 &       4400 & 1 &   0.88 &   0.19 &   -1.9 &       21.9 \\
    51A &   ATCDFS9 J0033201.55-274648.0 & 03:32:01.5 & -27:46:48.0 &       2501 &        21 &       3674 & 1 &   1.57 &   0.88 &   49.1 &       20.5 \\
    51B &   ATCDFS9 J0033201.29-274647.7 & 03:32:01.2 & -27:46:47.7 &       1302 &        21 &       2174 & 1 &   2.51 &   1.05 &   26.6 &       20.5 \\
     53 &   ATCDFS9 J0033200.84-273557.0 & 03:32:00.8 & -27:35:57.0 &       2648 &        33 &          0 & 0 &   0.00 &   0.00 &    0.0 &       20.1 \\
     54 &   ATCDFS9 J0033153.43-280221.2 & 03:31:53.4 & -28:02:21.2 &        962 &        26 &          0 & 0 &   0.00 &   0.00 &    0.0 &       18.7 \\
     55 &   ATCDFS9 J0033152.12-273926.5 & 03:31:52.1 & -27:39:26.5 &        355 &        31 &          0 & 0 &   0.00 &   0.00 &    0.0 &       21.4 \\
     56 &   ATCDFS9 J0033150.15-273948.1 & 03:31:50.1 & -27:39:48.1 &        176 &        29 &          0 & 0 &   0.00 &   0.00 &    0.0 &       19.7 \\
     57 &   ATCDFS9 J0033150.01-275807.0 & 03:31:50.0 & -27:58:07.0 &        125 &        19 &          0 & 0 &   0.00 &   0.00 &    0.0 &       20.0 \\
     58 &   ATCDFS9 J0033149.86-274839.2 & 03:31:49.8 & -27:48:39.2 &        658 &        19 &        861 & 1 &   1.29 &   0.92 &   23.8 &       19.1 \\
     59 &   ATCDFS9 J0033146.11-280026.5 & 03:31:46.1 & -28:00:26.5 &        337 &        21 &        392 & 1 &   1.68 &   0.41 &   13.8 &       20.9 \\
     60 &   ATCDFS9 J0033134.23-273828.3 & 03:31:34.2 & -27:38:28.3 &        124 &        33 &          0 & 0 &   0.00 &   0.00 &    0.0 &       20.8 \\
    61A &   ATCDFS9 J0033131.10-273815.8 & 03:31:31.1 & -27:38:15.8 &        955 &        56 &       1369 & 1 &   1.60 &   1.14 &    0.2 &       21.2 \\
    61B &   ATCDFS9 J0033130.96-273815.8 & 03:31:30.9 & -27:38:15.8 &        355 &        41 &        637 & 1 &   2.57 &   1.43 &  -10.2 &       21.2 \\
     62 &   ATCDFS9 J0033130.75-275735.3 & 03:31:30.7 & -27:57:35.3 &        133 &        25 &          0 & 0 &   0.00 &   0.00 &    0.0 &       20.7 \\
     63 &   ATCDFS9 J0033127.20-274247.3 & 03:31:27.2 & -27:42:47.3 &        337 &        20 &        427 & 1 &   2.48 &   0.52 &   -0.0 &       19.7 \\
     64 &   ATCDFS9 J0033127.04-275959.0 & 03:31:27.0 & -27:59:59.0 &         97 &        14 &          0 & 0 &   0.00 &   0.00 &    0.0 &       20.1 \\
     65 &   ATCDFS9 J0033124.90-275207.9 & 03:31:24.9 & -27:52:07.9 &       4095 &        21 &       8121 & 1 &   2.47 &   0.75 &   59.6 &       21.4 \\
     66 &   ATCDFS9 J0033123.30-274905.8 & 03:31:23.3 & -27:49:05.8 &        374 &        18 &          0 & 0 &   0.00 &   0.00 &    0.0 &       18.8 \\
     67 &   ATCDFS9 J0033117.35-280147.2 & 03:31:17.3 & -28:01:47.2 &        416 &        47 &          0 & 0 &   0.00 &   0.00 &    0.0 &       37.3 \\
    68A &   ATCDFS9 J0033117.04-275514.6 & 03:31:17.0 & -27:55:14.6 &        190 &        25 &        409 & 1 &   3.26 &   1.74 &    2.8 &       24.9 \\
    68B &   ATCDFS9 J0033115.05-275518.7 & 03:31:15.0 & -27:55:18.7 &       1900 &        33 &          0 & 0 &   0.00 &   0.00 &    0.0 &       22.9 \\
    68C &   ATCDFS9 J0033113.99-275519.2 & 03:31:13.9 & -27:55:19.2 &        152 &        20 &        478 & 1 &   4.84 &   1.85 &   30.9 &       26.0 \\
    68D &   ATCDFS9 J0033113.82-275518.9 & 03:31:13.8 & -27:55:18.9 &        112 &        34 &          0 & 0 &   0.00 &   0.00 &    0.0 &       27.6 \\
     70 &   ATCDFS9 J0033113.95-273910.8 & 03:31:13.9 & -27:39:10.8 &        511 &        23 &        595 & 1 &   1.61 &   0.52 &   -3.0 &       22.6 \\
\hline

\end{tabular}
}
\label{tab:catalog}
\end{table*}

\subsection{Completeness and Flux Density Bias}

\begin{figure*}
\includegraphics[width=8cm]{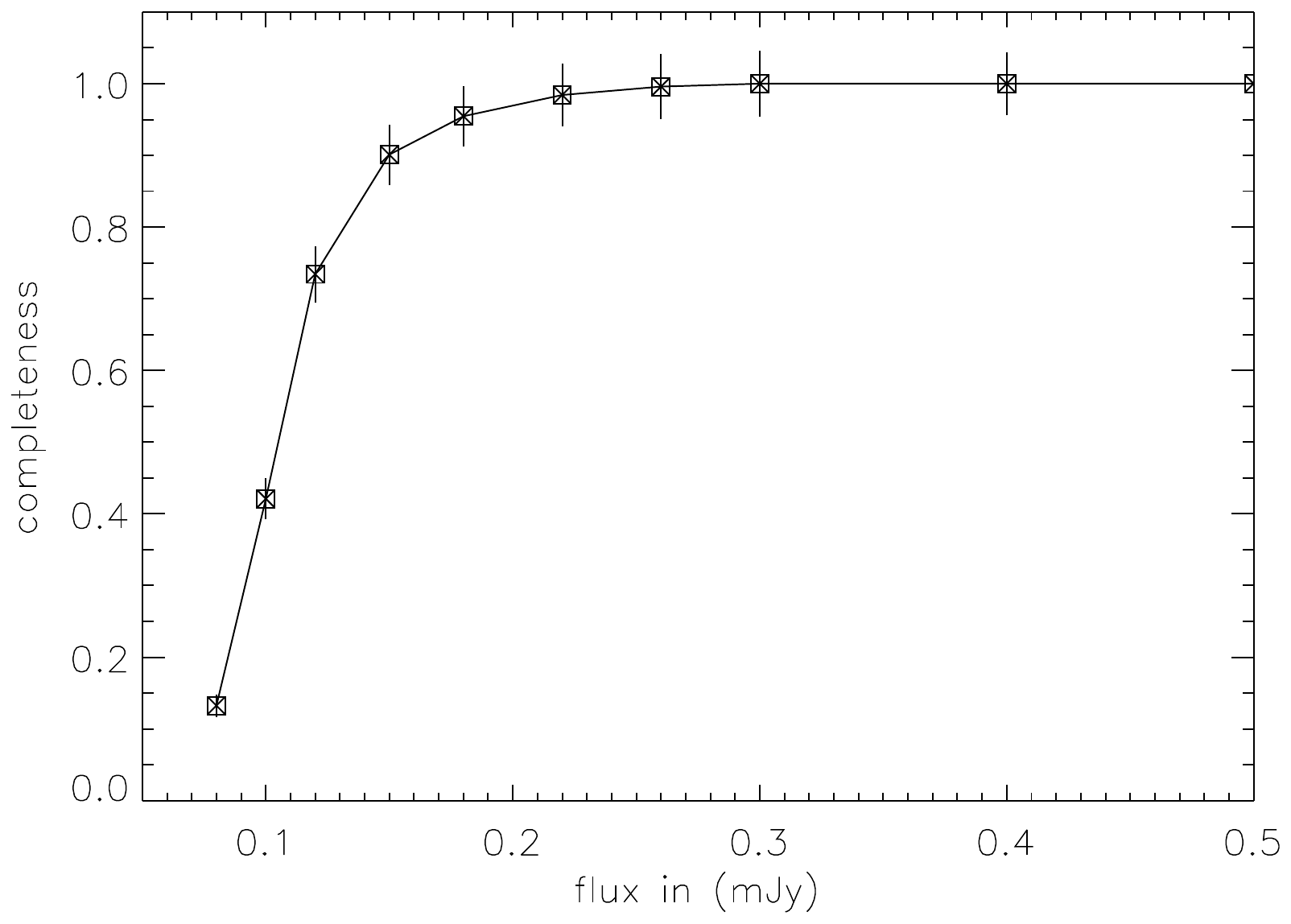}
\hspace{5mm}
\includegraphics[width=8cm]{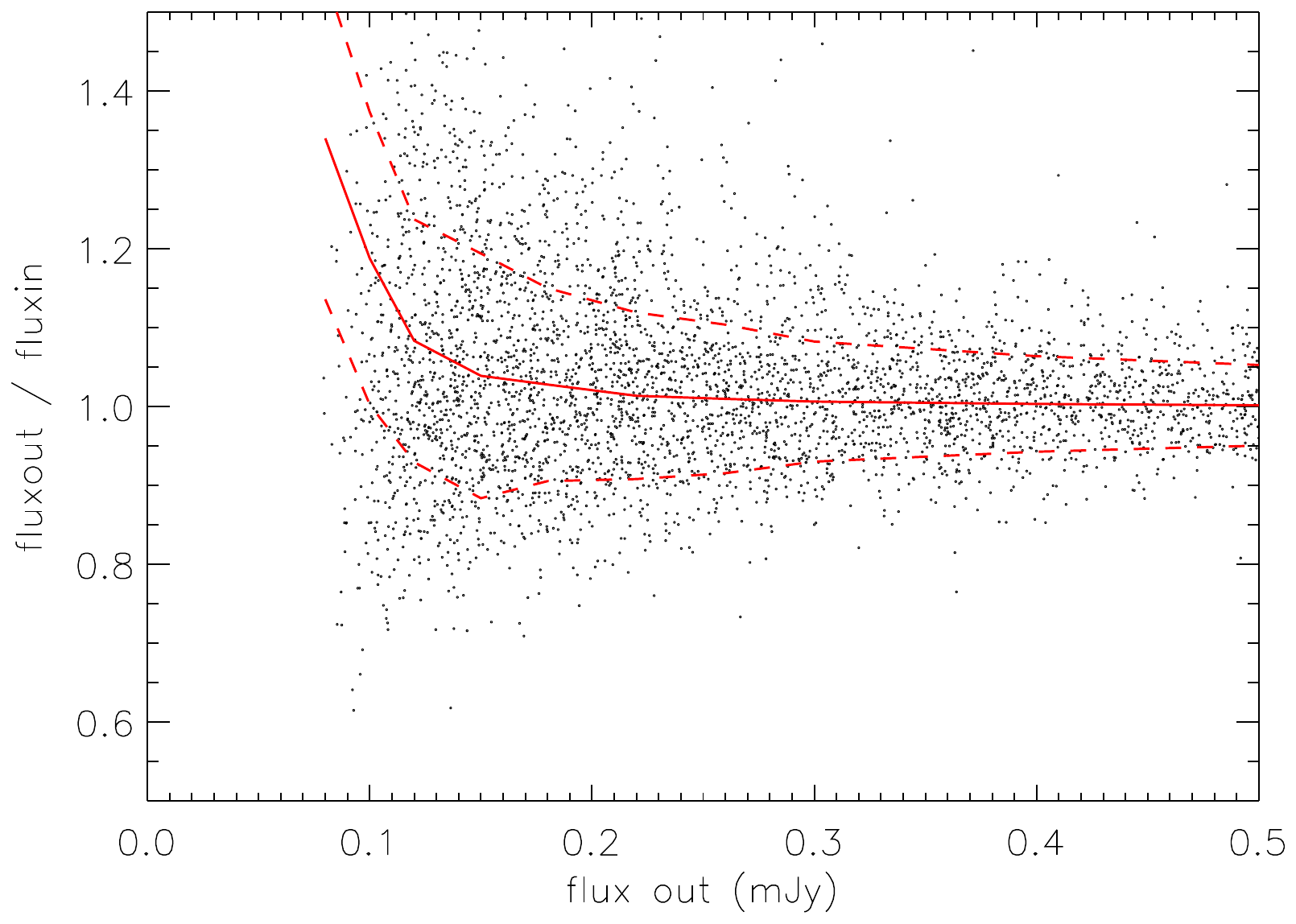}
\caption{LEFT: Completeness as a function of input flux density, as derived from the injected sources. Completeness is defined as the number of extracted sources divided by number of input sources. RIGHT: The distribution of output/input flux density as a function of output flux density for the injected sources. The solid red line is the median ratio from the simulation and the dashed lines mark the 1 sigma upper and lower bounds. The rapid upturn below about 0.12 mJy shows the effect of flux boosting at the faint end.} 
\label{im:sims}
\end{figure*}

For a simple S/N threshold source extractor, the completeness of the source catalogue as a function of flux density can be determined by the noise map variation. However, we used the MIRIAD task {\em sfind} which does not translate into a straightforward S/N threshold. 
As in H12 and H15, we performed Monte-Carlo simulations to determine the completeness of the source catalogue. The overall completeness level of the generated catalogue was determined by injecting sources over the full area of the mosaic and then extracting them with the same method that produced the catalogue, i.e. detection with {\em sfind}, measurement with {\em imfit}. 
Artificial sources were injected one at at time to avoid confusion effects, with input flux densities from 50 to 2500 $\mu$Jy. In total 10,000 artificial sources were injected for reliable statistics. The completeness is the fraction of input artificial sources which are recovered and extracted by {\em sfind}, and this is shown as a function of flux density in Figure \ref{im:sims}.  The completeness rises steeply from about 13\% at 80 $\mu$Jy to approximately 90\% at 150 $\mu$Jy. The 50\% completeness level occurs at approximately 105 $\mu$Jy. 

Faint sources that lie on a noise peak have increased flux densities and a higher probability of being detected than faint sources which lie on noise troughs. The average flux density for an ensemble of faint sources can therefore be increased due to noise, introducing a flux density bias. This effect is called 'flux boosting' and it is strongest in the faintest flux density bins. As in H12 and H15, the extent of flux boosting is estimated from the simulations, using the ratio of output to input flux density (Figure \ref{im:sims}). In the faintest bins we find that flux densities are boosted by about 34\% at 80 $\mu$Jy and 19\% at 100 $\mu$Jy, on average. Flux boosting is negligible for sources with flux densities brighter than about 150 $\mu$Jy.

\subsection{Source Sizes and Resolution Bias}

Faint extended sources may have peak flux densities that fall below the detection threshold, even though they may have large total flux densities. For source counts which are complete in terms of total flux density this so-called resolution bias must be determined.  We follow the technique of H12 and H15 to estimate the resolution bias. 

\begin{figure*}
\includegraphics[width=8cm]{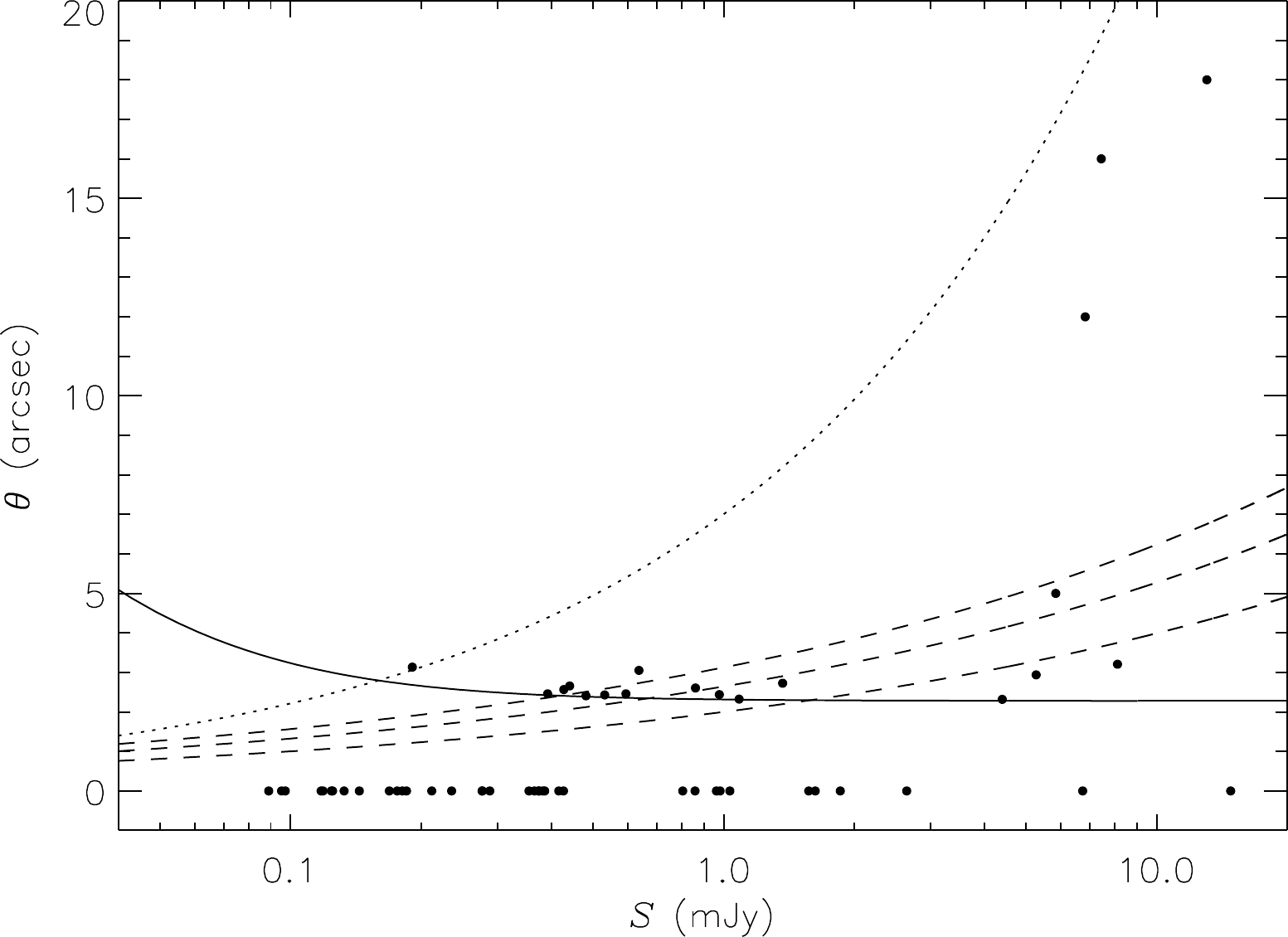}
\hspace{5mm}
\includegraphics[width=8cm]{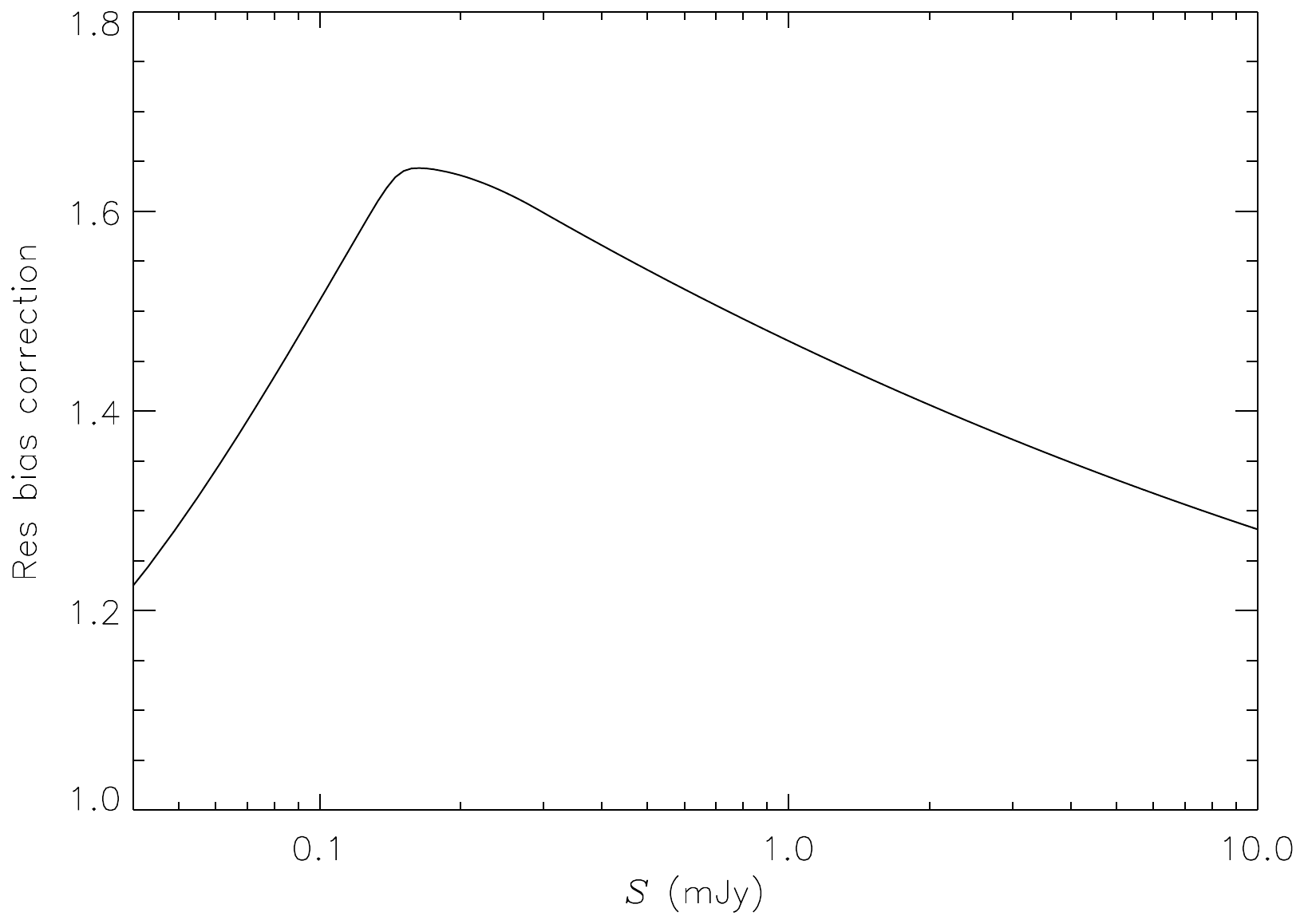}
\caption{LEFT: Fitted angular size as a function of total flux density. Unresolved or point sources are shown with an angular size of zero. Multiple component sources are plotted with the largest angular scale, or maximum distance between components. The solid line indicates the minimum angular size ($\theta_{\rm min}$) of sources in the survey, below which sources are considered unresolved. The dotted line shows the maximum angular size ($\theta_{\rm max}$) above which the survey becomes incomplete due to resolution bias. The dashed lines indicate the median source sizes expected from the Windhorst et al. (1990) relation, as a function of flux density, for a spectral index of $-0.8$, $-$0.5 and 0 (top to bottom) between 1.4 and 9.0 GHz. RIGHT: The resolution bias correction as a function of flux density, assuming the integral source size distribution from Windhorst et al. (1990).} 
\label{im:resbias}
\end{figure*}

In Figure \ref{im:resbias} we plot the angular sizes of the catalogued sources as a function of total flux density, where the angular size $\theta$ is defined as the geometric mean of the fitted Gaussian major and minor axes. Assuming a Gaussian beam, the maximum size ($\theta_{\rm max}$) a source of total flux density $S_{\rm tot}$ can have before it drops below the detection limit can be approximated by 
\begin{equation}
 S_{\rm tot}/\sigma_{\rm det} = \theta_{\rm max}^2 / ( b_{\rm min} b_{\rm maj} )\;, 
 \end{equation}
where $\sigma_{\rm det}$ is the detection limit and $b_{\rm maj}$ and $b_{\rm min}$ are the synthesized beam major and minor axes (FWHM), respectively.  Since the {\em sfind} detection limit varies across the image, we use the the pixels of the noise image to determine the relative weight of noise values ranging from 16.5 to 42.5 $\mu$Jy, assume a detection limit of 5$\sigma$ and calculate a noise-weighted $\theta_{\rm max}$. This is shown as a dotted line in Figure  \ref{im:resbias}. There is one extended source on the line and none above it, indicating that this maximum size estimate works well. 

An estimate of the minimum angular size ($\theta_{\rm min}$) a source can have is then estimated from Equation 1:
\begin{equation}
 S_{\rm int} / S_{\rm peak} = 1 + 10 / (S_{\rm peak} / \sigma)^{1.5} = \theta_{\rm min}^2 / (b_{\rm min} b_{\rm maj})
\end{equation}
Similar to $\theta_{\rm max}$ we calculate a noise-weighted $\theta_{\rm min}$ using the pixel distribution of the noise image.  The resulting $\theta_{\rm min}$ as a function of total flux density is plotted in Figure \ref{im:resbias} (solid line). The $\theta_{\rm min}$ constraint becomes important at low flux density levels, where $\theta_{\rm max}$ is smaller than a point source and therefore unphysical. 

Overplotted in Figure \ref{im:resbias} (dashed lines) is the median angular size relation for a 1.4 GHz sample from \cite{windhorst1990}: 
\begin{equation}
\theta_{\rm med} = 2\arcsec S_{\rm 1.4 GHz}^{0.30}
\end{equation}
where $S_{\rm 1.4 GHz}$ is in mJy. The 1.4 GHz flux densities were extrapolated to 9.0 GHz assuming a spectral index of 0, -0.5 and -0.8 between 1.4 and 9.0 GHz. At flux densities less than $\lesssim$0.5 mJy the radio sources are expected to have median sizes smaller than our beam, and this is consistent with our observations where we have only resolved a handful of sources at $S$ $\lesssim$ 0.5 mJy. 

The integral angular size distribution, $h(\theta)$, from \cite{windhorst1990} is: 
\begin{equation}
h(\theta) = \exp[-\ln 2 \, (\theta/\theta_{\rm med})^{0.62}]\;.
\end{equation}
This allows us to estimate the fraction of sources larger than the maximum detectable size, and hence missed by the survey. If we take the overall angular size limit to be $\theta_{\rm lim} = {\rm max}(\theta_{\rm max},\theta_{\rm min})$ then the resolution bias correction factor is 
\begin{equation}
c = \frac{1} {1 - h(\theta_{\rm lim})}\;.
\end{equation}
This correction factor is plotted in Figure \ref{im:resbias}. It has a maximum of about 1.7 at a flux density of 130 $\mu$Jy, where the limiting overall angular size, $\theta_{\rm lim}$, becomes dominated by $\theta_{\rm min}$. The strong caveat to the resolution bias correction is that the integral angular size distribution at these flux densities and this frequency is not well known. Sources are likely to be systematically smaller at 9 GHz compared to 1.4 GHz due to the lower sensitivity to steep-spectrum extended lobes. We notice that the 9 GHz sources brighter than 4 mJy have a median size of 3.2 arcsec, which lies below the Windhorst et al. (1990) distribution assuming a spectral index of $\alpha = 0$, but this result is from a small number of sources. We can not draw strong conclusions for flux densities fainter than $\sim$ 1 mJy but the point at which sources drop below our deconvolution limit is consistent with Windhorst et al. (1990) distribution. Resolution bias can be significant in ATCA 1.4 GHz observations (e.g. \citealp{huynh2005, hales2014}) and so it is likely to be even more important in the higher resolution 9 GHz observations. 

Another incompleteness comes from the lack of short baselines, which limits the maximum scale that a radio observation can reliably detect. With 6A and 6D ATCA configurations, our shortest baseline is 77m and there is reduced coverage for baselines shorter than 214m. We therefore start to be insensitive to extended structures $\gtrsim$ 30 arcsec and lose all sensitivity to extended structures $\gtrsim$ 90 arcsec. Only about one source in our survey area is expected to be greater than 30 arcsec in size, using the Windhorst et al. 1990 size distribution and radio source counts, and we don't expect a source larger than 90 arcsec in our small survey area. 

\section{Source Counts}

\label{sec:sourcecount}

The differential radio source counts, normalised to a non-evolving Euclidean universe ($dN/dS \; S^{2.5}$), were constructed from our catalogue. Integrated flux densities were used for resolved sources and peak (or point source) flux densities for the remainder. The multiple components of a single radio galaxy were summed and counted as a single source. The counts are summarised in Table \ref{tab:srccount} where for each bin we report the flux density interval, mean flux density, the number of sources detected ($N$), the number of sources after completeness and resolution bias corrections have been applied ($N_C$), and the normalised differential source count ($dN_{C}/dS \; S^{2.5}$ [Jy$^{1.5}$ sr$^{-1}$]). We estimate the error in the counts with $\sqrt{dN}/dS \; S^{2.5}$, approximately Poissonian, but with uncertainties in the completeness, flux boosting and resolution bias corrections added in quadrature. The completeness and flux boosting correction are estimated to have uncertainties of 5 to 7\% and 6 to 16\%, respectively. The uncertainty in the resolution bias is less clear: the Windhorst et al. 1990 work has few sources with flux densities less than a mJy and was performed at 1.4 GHz. We therefore take the conservative estimate that the resolution bias correction has a large 50\% uncertainty. 

\begin{table}
\centering
\caption{The 9.0 GHz source counts.}
\begin{tabular}{ccrrcc} \hline
$\Delta S$ & $\langle$S$\rangle$ & $N$ & $N_{C}$ & $dN_{C}/dS \, S^{2.5}$ \\
($\mu$Jy) & ($\mu$Jy) & & &  [Jy$^{1.5}$ sr$^{-1}$]  \\ \hline
80 -- 120 & 106 & 5 & 12.44 & 0.43 $\pm$ 0.17  \\
120 -- 175 & 150 & 4 & 7.23 & 0.43 $\pm$ 0.25 \\
175 -- 250 & 197 & 5 & 8.42 & 0.73 $\pm$ 0.41 \\
250 -- 370 & 321 & 4 & 6.37 & 1.16 $\pm$ 0.66 \\
370 -- 500 & 410 & 10 & 15.63 & 4.88 $\pm$ 1.92 \\
500 -- 700 & 563 & 2 & 3.06 & 1.37 $\pm$ 0.90 \\
700 -- 1000 & 894 & 5 & 7.41 & 7.02 $\pm$ 3.31 \\
1000 -- 1500 & 1125 & 3 & 4.38 & 4.42 $\pm$ 2.51 \\
1500 -- 2500 & 1766 & 4 & 5.67 & 8.86 $\pm$ 4.37  \\
2500 -- 4500 & 3216 & 3 & 4.11 & 14.3 $\pm$ 8.1  \\ 
4500 -- 6000 & 5558 & 2 & 2.64 & 48.4 $\pm$ 31.4 \\ \hline
\end{tabular}
\label{tab:srccount}
\end{table}

\begin{figure*}
\includegraphics[width=15cm]{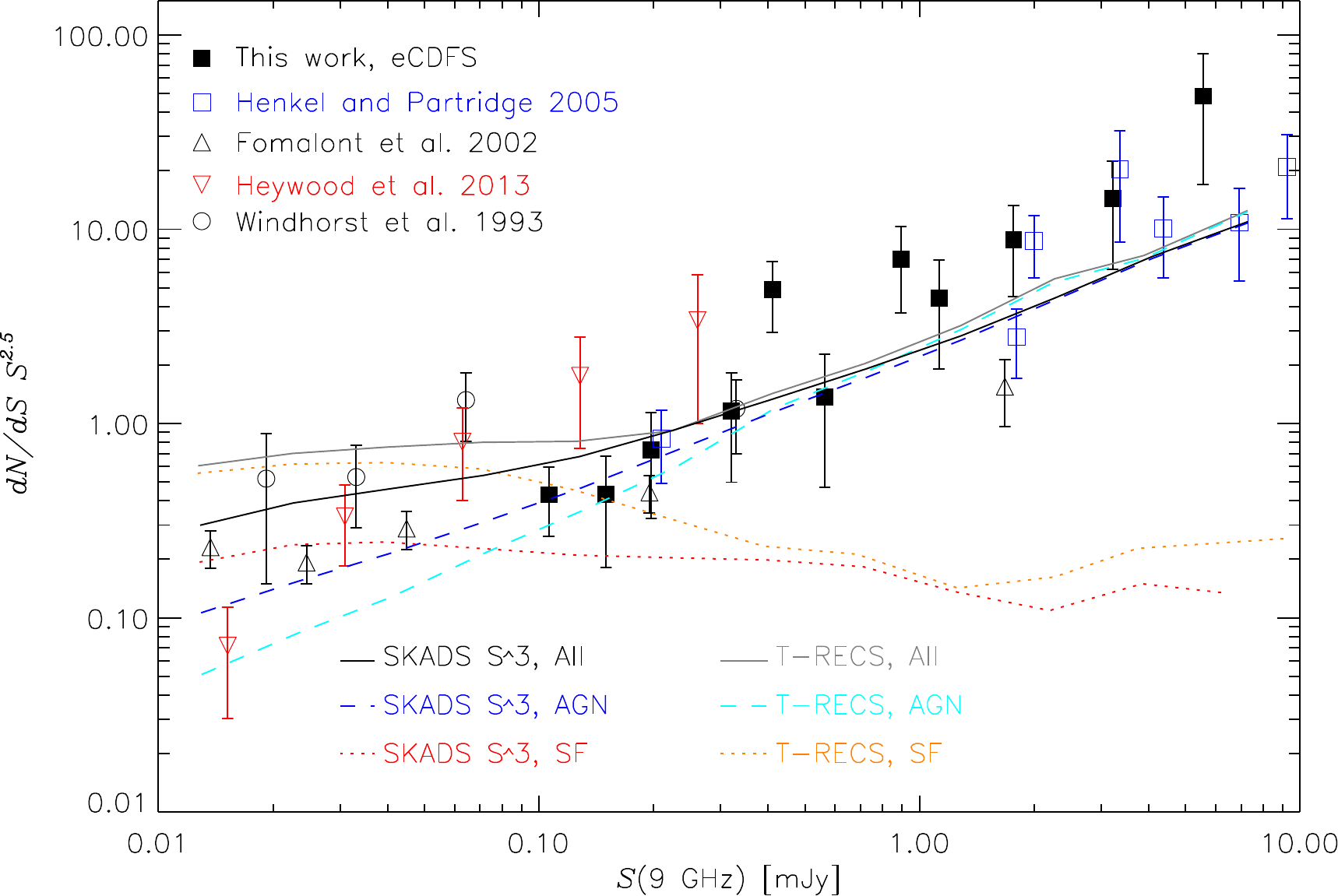}
\caption{Euclidean normalized 9 GHz differential source counts: this work (fill squares), Henkel and Patridge (blue squares); Fomalont et al. 2002 (black triangles); Heywood et al. 2013 (red upside-down triangles); Windhorst et al. (1993) (circles). Counts from this work are corrected for completeness, flux boosting and resolution bias. Vertical bars represent Poisson plus completeness, flux boosting, and resolution bias uncertainties. Also shown are model source counts from SKADS (Wilman et al. 2008) and T-RECS (Bonaldi et al. 2019) simulations.} 
\label{im:sourcecount}
\end{figure*}

Our source counts are shown in Figure \ref{im:sourcecount}, along with previous deep surveys performed on VLA at 8.4 GHz from \cite{windhorst1993,fomalont2002,henkel2005} and \cite{heywood2013}. 
These deep VLA surveys cover only small areas of sky consisting of several individual VLA pointings (Windhorst et al. 1993, Fomalant et al. 2002, Henkel and Partridge 2005), or a single deep pointing (Heywood et al. 2013a). Two large shallow mosaics, totalling more than a degree, were produced by Henkel and Partridge to cover the bright regime ($S > 1$ mJy). Our deep but wide ATCA mosaic provides the best statistics in the 0.1 to 1 mJy range. In comparing our source counts with VLA results we do not apply spectral index corrections, as these can be uncertain, but note that for a spectral index of $\alpha = -0.8$ the flux density difference is only 5\% between the VLA and our ATCA work (8.4 GHz vs 9.0 GHz). 

Source counts from the semi-empirical sky simulation developed for the SKA (SKADS S3-SEX, \citealp{wilman2008,wilman2010}) are shown in Figure \ref{im:sourcecount} as SKADS total counts (black solid line), SKADS AGN counts (blue dashed line) and SKADS star forming galaxy counts (red dotted line). These counts were linearly interpolated between the simulation frequencies of 4.86 and 18 GHz to the ATCA frequency of 9 GHz. A more recent simulation is the Tiered Radio Extragalactic Continuum Simulation (T-RECS, \citealp{bonaldi2019}), which is similar to SKADS but motivated by the need for an update to SKADS given a decade of more radio observations and improvements in modelling. T-RECS is based upon the evolutionary model of flat and steep spectrum AGN radio sources by \cite{massardi2010} (revised by \citealp{bonato2017}) and includes a star forming population with radio emission modelled following \cite{mancuso2015,bonato2017} and \cite{cai2013}. We use the medium simulation of T-RECs consisting of AGN and SF galaxies over 25 square degrees and show source counts from the simulation's 9.2 GHz flux densities in Figure \ref{im:sourcecount}. 

At bright flux densities ($S > 1$mJy) our source counts are consistent with Henkel and Partridge 2005.  Although there are large uncertainties, the observed counts are in general well represented by the SKADS and T-RECS models, which show little difference for $S \gtrsim 0.3$ mJy. The observed counts for $S \gtrsim 0.5$ mJy appear to be greater than the simulations but the discrepancy is only about 0.3 to 0.5 dex, or roughly a factor of two, and generally within 1 sigma of the count uncertainty. The bump in the observed counts in the 0.37 to 0.5 mJy and 0.7 to 1 mJy bins is likely due to large scale structure and cosmic variance. We find that 4/10 and 2/5 of the sources in the 0.37 to 0.5 mJy and 700 to 1 mJy bins lie at redshifts associated with known large galaxy clusters in the eCDFS (see next Section \ref{sec:agnsf} and discussion of redshifts). 

At low flux densities ($S \lesssim 0.3$ mJy) the observed counts show a spread. This is a well known issue for radio source counts at the lowest flux densities. Cosmic variance can cause considerable scatter due to the small size of the fields in deep surveys. Survey systematics introduced by calibration, deconvolution and source extraction algorithms, and corrections applied to the raw source counts also tend to be largest at the faint end, adding to the uncertainty. For example, the deep VLA surveys consist of a single, or several, individual primary-beam areas with non-uniform sensitivity, so these counts contain large corrections for primary beam gain. The resolution bias of our counts is greatest in our faintest bins. A clustering analysis of the SKADS simulations concluded that cosmic variance is the major cause of the difference in radio source counts from single/several pointing surveys at $S_{\rm 1.4GHz} \gtrsim 0.1$ mJy \citep{heywood2013b}, equivalent to  $S_{\rm 9GHz} \gtrsim 0.025$ mJy. Nevertheless the faintest counts do follow the simulation of SKADS and T-RECS in general (solid black and grey lines), with the exception of the faintest bin from Heywood et al. 2013. 

The T-RECS simulation diverges from the SKADS count for $S \lesssim 0.1$ mJy, and this appears to be due to the larger star forming population in the T-RECS simulation. We have detected only one star forming galaxy out of 55 galaxies in our 9 GHz sample (see Section \ref{sec:agnsf}), suggesting that our ATCA eCDFS survey does not yet probe the regime where star forming galaxies become significant. This is consistent with the SKADS simulations but the T-RECS simulation suggests $\sim$50\% of the sources in our faintest two bins will be star forming galaxies, which appears to be an overprediction. Higher frequency surveys are biased to flatter spectrum sources, i.e. AGN, but this can not completely explain the low number of star forming galaxies in our observations. The difference may be due to how the star forming galaxies in T-RECS are modelled or because radio-quiet AGN, whose radio emission is not dominated by AGN, are not accounted for in the T-RECS simulation. However, our observations are not sensitive to typical local star forming spirals and we find this can account for most of the difference between our star forming galaxy numbers and those in the T-RECS simulation (see Section \ref{sec:agnsf}).

\section{Nature of Faint 9 GHz Radio Population}

\subsection{Star Forming Galaxies vs Active Galactic Nuclei} 

\label{sec:agnsf}

We investigate whether the 9 GHz radio sources are dominated by star forming or active galactic nuclei processes. In the eCDFS, VLA 1.4 GHz radio sources were identified with an optical/IR counterpart by \cite{bonzini2013}. Bonzini et al. 2013 separate radio sources into radio loud AGN (RL AGN), radio quiet AGN, and star forming galaxies using the multi-band data of this well-studied field. Firstly, they identify radio loud AGN using a method based on the observed 24$\mu$m to 1.4 GHz flux density ratio $q_{24 obs} = \log ( S_{24 \mu m} / S_{\rm 1.4 GHz})$. Sources are classified as radio loud if they lie more than 2$\sigma$ below the $q_{24 obs}$ ratio expected of a typical star forming galaxy SED (M82). Radio sources that lie above the $q_{24 obs}$ threshold were then classified by Bonzini et al. as radio quiet if they showed clear signs of AGN activity in X-Ray (XRay luminosity $> 10^{42}$ erg s$^{-1}$) or MIR colour colour space (\citealp{donley2012} colour wedge). The remaining sources were classified as star forming galaxies. These simple criteria do not account for composite AGN/SF sources or optically-selected Seyferts which have a $q_{24 obs}$ ratio expected of SF galaxies. In other words, this criteria may misclassify some radio-quiet AGN as SF galaxies, but we have very few SF galaxies in our sample.

We find Bonzini et al. matches for 52/55 (95\%) of the radio galaxies in our 9 GHz sample.  
We examined the three 9 GHz sources with no Bonzini et al. identification in detail:

{\em ID 43} is a faint inverted spectrum source which is detected at 5.5 GHz and 9.5 GHz, but not detected in VLA 1.4 GHz imaging. It has a 5.5 GHz flux density of 92 $\pm$ 11 $\mu$Jy (H15) and 9.0 GHz flux density of 144 $\pm$ 29$\mu$Jy from this work, and hence a spectral index of $\alpha = 0.91 \pm 0.47$. It has a faint HST F606W band magnitude of 28.37 \citep{giavalisco2004}. The photometric redshift for this galaxy has been determined by several teams to range from 2.30 to 2.531 \citep{wuyts2008,bundy2009,rafferty2011,hsu2014}, which is consistent given differences in data and SED templates used. We take the photometric redshift of $z$ = 2.4174 from Hsu et al. 2014 as it uses the latest multiwavelength catalogue from the CANDELS survey \citep{guo2013}. This source also has a Chandra detection, and given $z$ = 2.4174 its XRay (2 - 10 keV) luminosity is $10^{42.877}$ erg s$^{-1}$, so this source is an XRay AGN. If we extrapolate the radio SED then the 1.4 GHz flux density is expected to be $\sim$26 $\mu$Jy, consistent with the non-detection in the VLA 1.4 GHz imaging (4$\sigma$ limit of $\sim$30 $\mu$Jy). Given no Spitzer MIPS 24 $\mu$m detection and a $S_{24 obs}$ limit of 30 $\mu$Jy from \cite{magnelli2009} then this source is expected to have $q_{24 obs} < 0.0$. This ratio is borderline between radio-loud and radio-quiet (see Bonzini et al. 2013 Figure 2), however we class it as a radio-quiet AGN given it is so faint in the radio bands.

{\em ID 45} is not detected in Spitzer IRAC or MIPS imaging. It has a very faint F140W detection in CANDELS HST imaging \citep{skelton2014}, with an AB magnitude of 24.47.  This radio galaxy is bright and resolved at 1.4 GHz , with an integrated flux density of 1.38 $\pm$ 27 mJy (Miller et al. 2013). It was also resolved at 5.5 GHz with a flux density of 424 $\pm$ 17.8 $\mu$Jy (H15).  The radio spectral index between 1.4 and 5.5 GHz is $\alpha = -0.86 \pm 0.05$.  The source has an integrated flux density of about 191 $\mu$Jy at 9 GHz and measured spectral index of  $\alpha = -1.62 \pm 0.47$ between 5.5 and 9.0 GHz. Hence this radio source appears to have a steepening spectrum but the 9 GHz imaging is likely to be resolving out flux. The MIPS 24 $\mu$m limit is 30 $\mu$Jy (Magnelli et al. 2009) so this source has $q_{24 obs} < 1.7$, indicating it is a radio-loud AGN.  Skelton et al. 2014 estimate the photometric redshift of this source to be $z$ = 5.7755, and at that redshift the source would have an XRay (2 - 10 keV) luminosity of 7.1 $\times$ 10$^{43}$ erg s$^{-1}$, consistent with an AGN \citep{luo2017}. The IRAC 3.6 $\mu$m detection limit is approximately 0.15 $\mu$Jy (3$\sigma$), so this source meets the \cite{zinn2011} criteria for Infrared Faint Radio Source (IFRS), and is consistent with the paradigm that these sources are high redshift radio loud AGN. If the photometric redshift is correct, then this is highest redshift IFRS currently known (see \citealp{orenstein2019}).

{\em ID 61} is the eastern lobe of a wide-angle-tailed radio-loud galaxy detected by ATCA and VLA imaging at 1.4 GHz \citep{norris2006,miller2013}, and ATCA imaging at 5.5 GHz (H15). Identification of the optical counterpart is made difficult by a bright star just 20 arcsec away to the south. However, a faint counterpart can be found in COMBO17 imaging with an R band AB magnitude of 25.3 \citep{wolf2004} and photometric redshift of $z$ = 1.16 (68\% lower and upper confidence levels of $z$ = 0.88 and $z$ = 1.32) (Rafferty et al. 2011). 

From our analysis of these three radio sources and Bonzini et al. 2013 classifications for the remainder, we find 50/55 (91\%) of our 9 GHz sources are radio loud AGN, 4/55 (7\%) are radio quiet AGN, and only one source (2\%) is a star forming galaxy (see Table \ref{tab:cpart}). This is very similar to the \cite{whittam2015} sample of radio sources (10C, $S_{\rm 15.7 GHz} > 0.5$ mJy), who found 90/96 (94\%) are radio loud, 1/96 (1\%) is definitely radio quiet, and a further 5/96 (5\%) are borderline between radio loud and radio quiet. 

Much deeper 9 GHz surveys are required to probe the star forming population significantly, and the high resolution of the 9 GHz imaging may be resolving out nearby star forming spirals.  The brightness temperature sensitivity is given by: 
\begin{equation}
 T_b = \frac{S_\nu}{\Omega_{beam}} \frac{c^2}{2 k \nu^2} \; , 
 \end{equation}
where is $\nu$ the observing frequency, $S_\nu$ is the integrated flux density, and $\Omega_{beam}$ is the beam solid angle. The constants $c$ and $k$ are the speed of light and the Boltzmann constant, respectively. Taking the nominal $5\sigma$ limit of 100 $\mu$Jy and image beam parameters of 4.0 and 1.3 arcsec, we find a 9 GHz brightness temperature limit of 0.29 K, or equivalently, a surface brightness limit of 0.017 mJy/arcsec$^2$. This is the limit of bright face-on spirals such as NGC 253 (see Figure 2 of \citealp{condon1991}), hence our observations are only sensitive to the brightest local spirals and starbursts. 

To estimate the number of missing star forming galaxies from our sample we use the VLA 1.4 GHz radio sources in the eCDFS from Bonzini et al. (2013) which are classified as star forming galaxies and predict their 9 GHz radio flux densities assuming a synchrotron spectral index ($\alpha = -0.7$). We find 5 star forming galaxies with a predicted 9 GHz flux density greater than 89 microJy (our faintest detected source), and all are local, with $0.08 < z < 0.15$. Only 1/5 of these were detected by our observations (ID 11). The nearest of the missing star forming galaxies has a relatively large predicted 9 GHz flux density of 240 $\mu$Jy, but its peak flux density was just below the threshold for detection (i.e. it is bright but large and extended). The remaining 3 undetected 1.4 GHz star forming sources have predicted 9 GHz flux densities of 102 to 104 $\mu$Jy. The star forming fraction in the faintest two bins would increase from 1/9 (11\%) to 5/13 (38\%) if these were detected by our observations. Thus local missing star forming galaxies can account for most of the discrepancy between our star forming galaxy counts and those from T-RECS.

Using the wealth of multiwavelength data available in the eCDFS, we were able to assign a reliable literature spectroscopic redshift to 41/55 ($\sim$75\% ) of the radio sources, and a photometric redshift to 13 of the 14 remaining sources. The radio source without a spectroscopic or photometric redshift is at the edge of the radio image and lies just west of the 30 $\times$ 30 arcmin eCDFS region with the best multiwavelength coverage. The redshifts and resulting radio luminosity of the radio sources are summarised in Table \ref{tab:cpart}. 

\begin{figure}
\includegraphics[width=8cm]{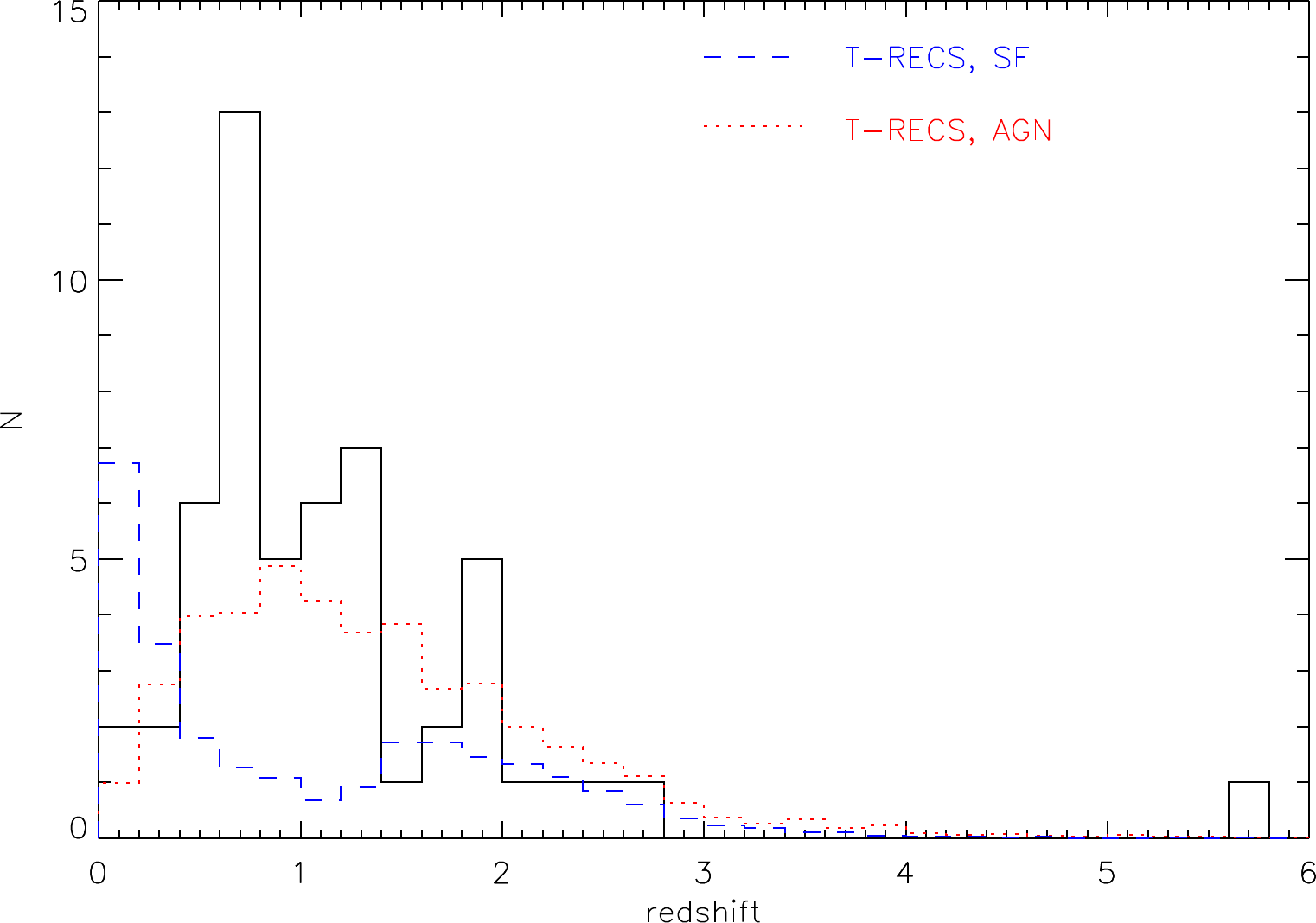}
\caption{The redshift distribution of our 9 GHz radio sample. For comparison the redshift distributions for AGN and SF galaxies in the T-RECS simulation, given our flux density cut and survey area, are shown as red-dotted and blue-dashed lines, respectively.} 
\label{im:redshift}
\end{figure}

In Figure \ref{im:redshift} we plot the observed redshift distribution of the 9 GHz radio sources. There is a peak in the observed distribution at $z$ $\sim$ 0.6 -- 0.7, which corresponds to known galaxy overdensities at $z \simeq 0.68$ and $z \simeq 0.73$ \citep{gilli2003,dehghan2014}. The redshift distribution for all AGN sources in the T-RECS simulation is over-plotted for comparison. The full T-RECS AGN model redshift distribution was normalised to our flux cut and area. The small numbers in each bin makes it difficult to draw definitive conclusions, but the broad shape of the model AGN redshift distribution is consistent with the observed distribution,  taking into account the known large scale structure at $z$ $\sim$ 0.7. In Figure \ref{im:redshift} we also show the T-RECs redshift distribution for star forming galaxies, again normalised for our flux cut and area coverage, and find that there are a significant number of star forming galaxies in the lowest redshift bin of the simulation which are not in our observations. This is consistent with our analysis above that our observations are missing local star forming galaxies due to the brightness temperature limit. However, T-RECS also predicts a low number of star forming galaxies up to redshift $z \sim 2.5$, on the level of 1 to 2 in each bin, which are not reflected in our observations. 

\begin{figure}
\includegraphics[width=8cm]{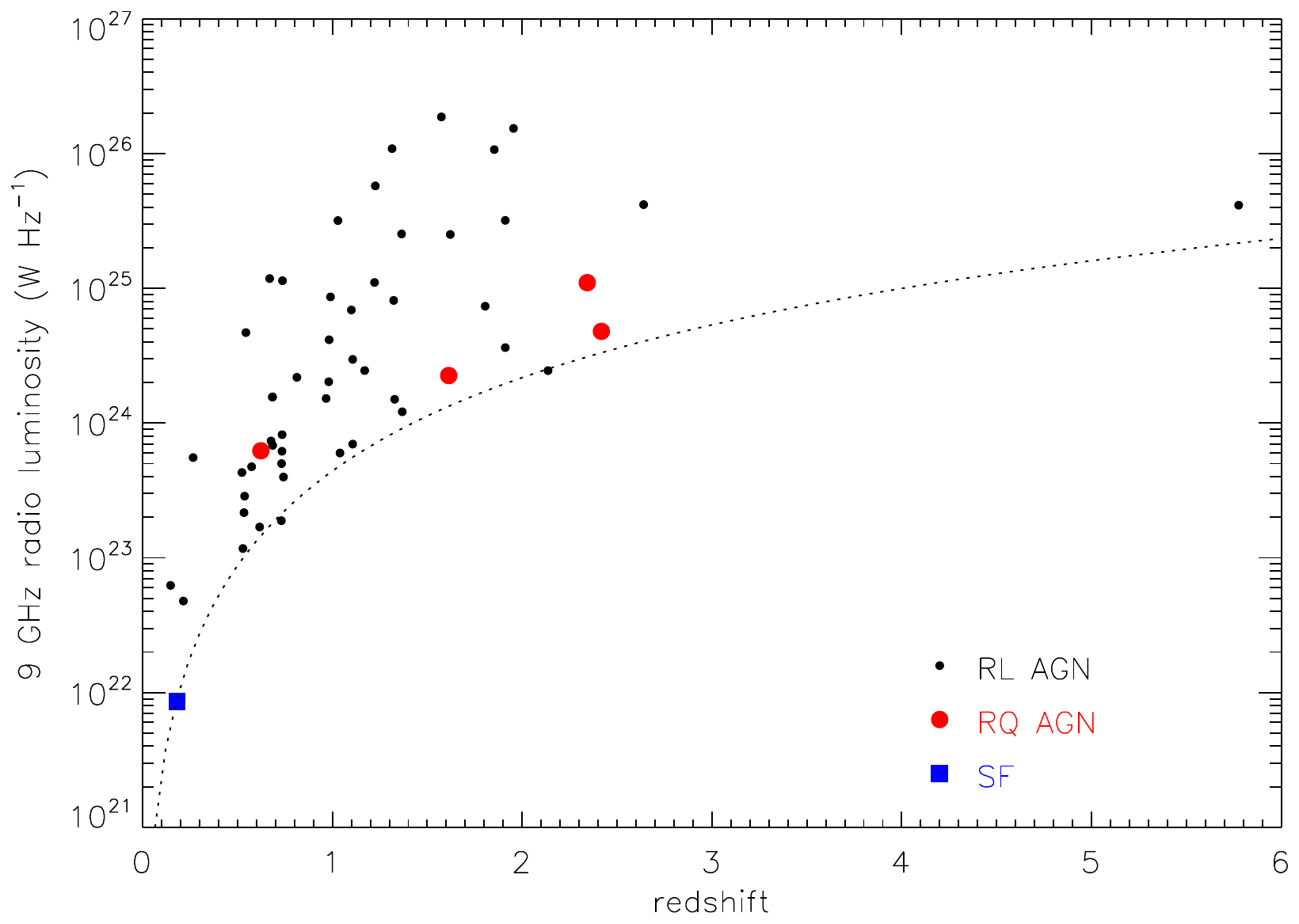}
\caption{Radio luminosity as a function of redshift for the 9 GHz radio sources. Radio-loud AGN are shown as small black circles, radio-quiet AGN are marked by larger red circles, and the one star forming galaxy is marked by the blue square. The dotted line is the approximate detection limit of the survey.} 
\label{im:radiolum}
\end{figure}

Figure \ref{im:radiolum} shows the 9 GHz monochromatic radio luminosity (in units of W Hz$^{-1}$) versus redshift for the radio sources, for the radio-loud, radio quiet and SF classes. The dotted line in Figure  \ref{im:radiolum} denotes the detection limit of our 9 GHz survey using the approximate 5 $\sigma$ limit of 100 $\mu$Jy. As expected the one star forming galaxy is at low redshift and is low radio luminosity ($P_{\rm 9 GHz}$ = 8.6 $\times$ 10$^{21}$ W Hz$^{-1}$. The 1.4 GHz luminosity of this source (using the VLA detection) is $P_{\rm 1.4 GHz}$ = 3.7 $\times$ 10$^{22}$ W Hz$^{-1}$, which corresponds to a star formation rate of $\sim$ 23 M$_\odot$ yr$^{-1}$ using the calibration of \cite{kennicutt2012}, so this is only a moderate starburst. The radio quiet AGN are detected across a wide redshift range, and tend to be at the lower end of radio luminosities for their particular redshift bin.  Radio loud AGN are found across the full radio luminosity range.

\begin{table*}
\caption{Galaxy classification and redshift information for the 9 GHz radio sources.}
\begin{tabular}{llclll} \hline
ID & classification & 9 GHz radio luminosity & redshift & redshift type & redshift reference \\
    &               & log (W Hz$^{-1}$) &          &               &                     \\ \hline
1 & RL AGN & 23.83 & 0.6852 & spectroscopic & \cite{mao2012} \\
2 & RL AGN & 24.39 & 1.17 & photometric & Bonzini et al. 2013 \\
3 & RL AGN & 23.63 & 0.5233 & spectroscopic & Mao et al. 2012 \\
5 & RL AGN & 23.46 & 0.53739 & spectroscopic & Cooper et al. 2012  \\
6 & RL AGN & 24.19 & 0.68466 & spectroscopic & Cooper et al. 2012 \\
7 & RL AGN & 25.62 & 2.64 & photometric & \cite{rafferty2011} \\
8 & RL AGN & 24.62 & 0.98268 & spectroscopic & Cooper et al. 2012 \\
9 & RL AGN & 24.34 & 0.81281 & spectroscopic & \cite{childress2017} \\
10 & RL AGN & 25.50 & 1.0291 & spectroscopic & \cite{balestra2010} \\
11 & SF & 21.93 & 0.18107 & spectroscopic & Cooper et al. 2012 \\
12 & RL AGN & 23.70 & 0.732 & spectroscopic & Silverman et al. 2010 \\
13 & RQ AGN & 23.79 & 0.6229 & spectroscopic & Cooper et al. 2012 \\
14 & RL AGN & 25.07 & 0.66931 & spectroscopic & Cooper et al. 2012 \\
19 & RL AGN & 22.80 & 0.14726 & spectroscopic & Childress et al. 2017 \\
20 & RL AGN & 24.47 & 1.107 & photometric & Rafferty et al. 2011 \\
21 & RL AGN & 23.34 & 0.5342 & spectroscopic & Cooper et al. 2012 \\
22 & RL AGN & 24.31 & 0.9807 & spectroscopic & \cite{lefevre2004}\\
23 & RL AGN & 25.04 & 1.222 & spectroscopic & \cite{szokoly2004} \\
25 & RL AGN & 26.04 & 1.3145 & photometric & \cite{hsu2014}\\
29 & RL AGN & 26.19 & 1.95429 & spectroscopic & Childress et al. 2017 \\
33 & RQ AGN & 24.35 & 1.613 & spectroscopic & \cite{vanzella2008}\\
34 & RL AGN & 22.68 & 0.2149 & spectroscopic & Cooper et al. 2012 \\
35 & RL AGN & 25.06 & 0.73703 & spectroscopic & Childress et al. 2017 \\
36 & RL AGN & 23.23 & 0.61651 & spectroscopic & Childress et al. 2017 \\
37 & RL AGN & 23.07 & 0.528547 & spectroscopic & Cooper et al. 2012 \\
38 & RL AGN & 24.84 & 1.10 & spectroscopic & Silverman et al. 2010 \\
40 & RL AGN & 23.79 & 0.734322 & spectroscopic & Cooper et al. 2012 \\
41 & RL AGN & 25.50 & 1.911 & photometric & Rafferty et al. 2011 \\
42 & RL AGN & 23.60 & 0.74208 & spectroscopic & Cooper et al. 2012 \\
43 & RQ AGN & 24.68 & 2.4172 & photometric & Hsu et al. 2014 \\
44 & RL AGN & 23.28 & 0.731 & spectroscopic & Silverman et al. 2010 \\
45 & RL AGN & 25.62 & 5.7755 & photometric & Skelton et al. 2014 \\
46 & RL AGN & 26.27 & 1.5743 & spectroscopic & Cooper et al. 2012 \\
47 & RL AGN & 24.56 & 1.911 & photometric & Rafferty et al. 2011 \\
48 & RL AGN & 25.40 & 1.6212 & spectroscopic & Cooper et al. 2012 \\
49 & RL AGN & 23.91 & 0.73504 & spectroscopic & Cooper et al. 2012 \\
50 & RL AGN & 24.67 & 0.54411 & spectroscopic & Childress et al. 2017 \\
51 & RL AGN & 26.03 & 1.853 & photometric & Cowley et al. 2016 \\
53 & RL AGN & 23.74 & 0.266 & spectroscopic & Silverman et al. 2010 \\
54 & RL AGN & 24.91 & 1.32293 & spectroscopic & Childress et al. 2017 \\
55 & RQ AGN & 25.04 & 2.3428 & spectroscopic & Danielson et al. 2017 \\
56 & RL AGN & 24.18 & 1.328 & photometric & Rafferty et al. 2011 \\
57 & RL AGN & 23.84 & 1.107 &  photometric & Rafferty et al. 2011 \\
58 & RL AGN & 24.19 & 0.6834 & spectroscopic & Eales et al. 2009 \\
59 & RL AGN & 23.68 & 0.57443 & spectroscopic & Childress et al. 2017 \\
60 & RL AGN & 23.78 & 1.040 & spectroscopic & Silverman et al. 2010 \\
61 & RL AGN & 24.94 & 0.99 & photometric & Wolf et al. 2008 \\
62 & RL AGN & 24.08 & 1.3681 & spectroscopic & Danielson et al. 2017 \\
63 & RL AGN & 24.87 & 1.80514 & spectroscopic & Childress et al. 2017 \\
64 & RL AGN & 24.39 & 2.136 & photometric & Rafferty et al. 2011 \\
65 & RL AGN & 25.76 & 1.2262 & spectroscopic & Danielson et al. 2017 \\
66 & RL AGN & 24.18 & 0.96685 & spectroscopic & Cooper et al. 2012 \\
67 & RL AGN & 23.87 & 0.6767 & spectroscopic & Norris et al. 2006 \\
68 & RL AGN & 25.40 & 1.36458 & spectroscopic & Cooper et al. 2012 \\
70 & RL AGN & --- & --- & --- & --- \\
\hline
\end{tabular}
\label{tab:cpart}
\end{table*}

\subsection{Radio Spectral Energy Distributions}

The existing radio observations of the eCDFS allows us to study the broadband radio spectral energy distributions of faint radio sources. The VLA observations reach 7.4 $\mu$Jy/beam rms for a 2.8 $ \times$ 1.6 arcsec beam at 1.4 GHz (Miller et al. 2013), and the ATCA observations achieve 8.6 $\mu$Jy/beam rms, for a beam of 5.0 $\times$ 2.0 arcsec (H15).  The radio spectral energy distributions (SED) and spectral indices of the 5.5 GHz-selected faint radio population in eCDFS were presented in our previous work (H15). The conclusion was that the deep 5.5 GHz observations are starting to probe the star forming population but a significant fraction (39\%) of even the faintest sources show a flat or inverted radio spectral index, and several radio sources showed deviation from a log-linear fit in their radio SED consistent with steepening or a gigahertz-peaked spectrum. 

\begin{figure}
\includegraphics[width=8cm]{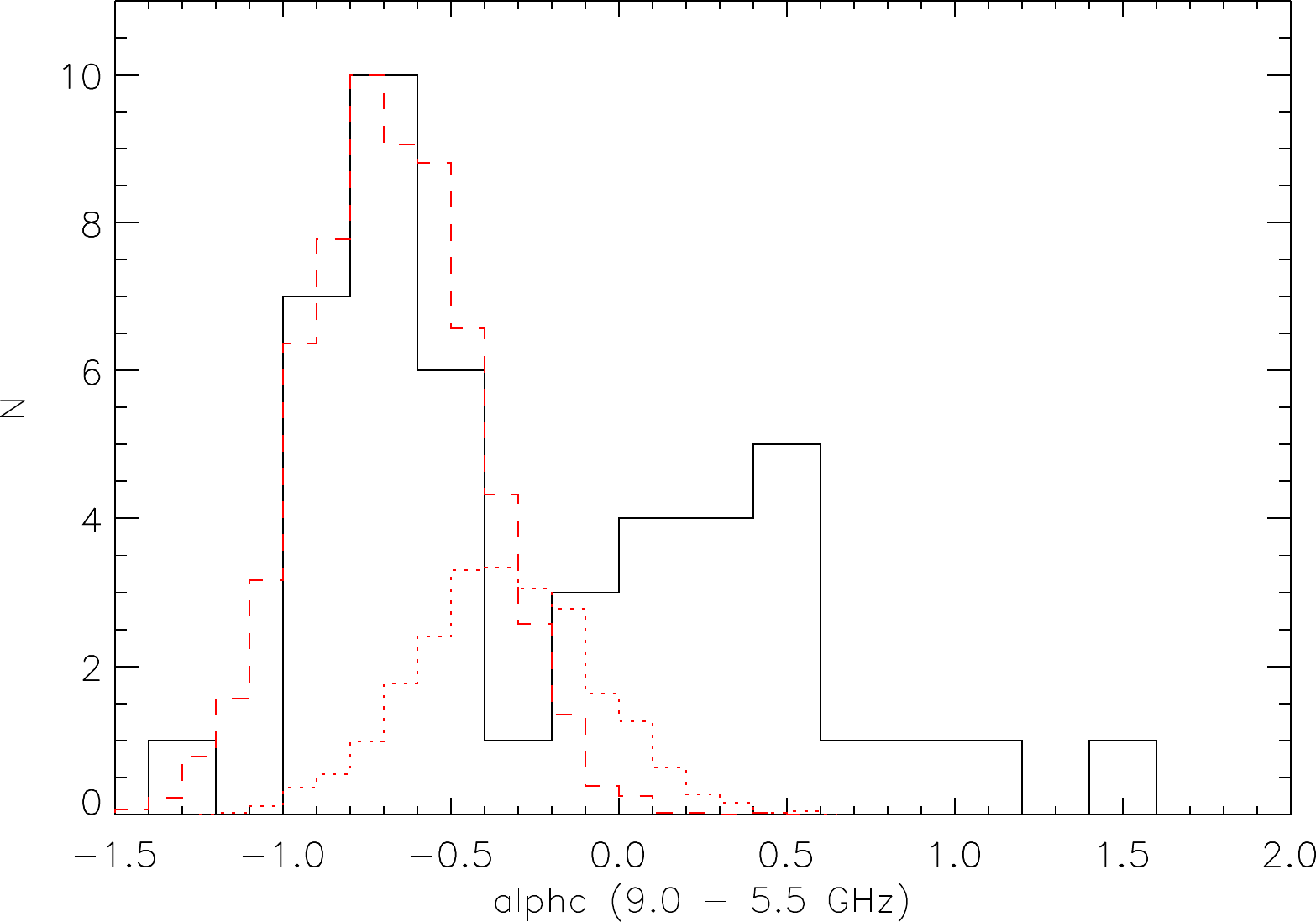}
\caption{The distribution of the spectral index between 5.5 and 9.0 GHz for our 9 GHz sample. Nine steep spectrum ($\alpha < -1.0$) sources are not included as they are missing 9 GHz flux. The T-RECS model steep spectrum (dashed red line) and flat spectrum AGN sources (dotted red line) are plotted for comparison. } 
\label{im:alphahist}
\end{figure}

To study the spectral index of the 9 GHz selected sample we matched the 9 GHz sources to the Miller et al. 2013 VLA 1.4 GHz survey of the eCDFS and the 5.5 GHz survey from our earlier ATCA work (H15). Multiple component sources were combined so that the the total flux densities were used in calculating the spectral index. The resulting spectral indices are collated in Table \ref{tab:alphas}.  We examined the sources with the steepest spectral indices between 5.5 and 9.0 GHz ($\alpha < -1.0$) which have a more canonical spectral index at 1.4 to 5.5 GHz ($\alpha$ closer to the synchrotron value of $-0.7$). We find ten of these sources are extended at 1.4 and/or 5.5 GHz, but are point-like or have missing components at 9 GHz, suggesting they are missing some 9 GHz flux density or have non-detected lobes due to the higher resolution and lower sensitivity of the 9 GHz imaging. Removing these ten sources with spuriously steep spectra results in a spectral index $\alpha_{5.5}^{9.0}$ distribution that is almost bimodal: the distribution shows a population with a canonical synchrotron spectral index peaking at close to $-0.7$ and a second population of flat and inverted spectrum sources with $\alpha >  -0.3$ (Figure \ref{im:alphahist}). We find 20/55 or 36\% of the 9 GHz population are flat or inverted with $\alpha >  -0.3$. The fraction of flat spectrum sources in AGN models of the the T-RECS simulation (Bonaldi et al. 2019), after applying a flux density cut of 50 $\mu$Jy at 5.0 GHz and 100 $\mu$Jy at 9.2 GHz to reproduce the survey selection criteria, is $\sim$17\% suggesting that their model is missing a significant number of flat spectrum AGN at 9 GHz. 

The spectral index distribution for flat and steep spectrum AGN sources in the T-RECS simulation with our selection criteria ($S_{\rm 5 GHz} < 50 \mu$Jy and $S_{\rm 9.2 GHz} < 100 \mu$Jy) is shown in Figure \ref{im:alphahist} for comparison, re-normalised to the peak of our observed distribution. The steep AGN component appears to be reflected in our observed spectral index distribution, however the model flat spectrum component has a peak at about $\alpha = -0.5$. The observed inverted spectrum sources ($\alpha > 0$) appear to be missing from the simulations and seem to be not well-handled by the radio population models of T-RECS. 

\begin{figure}
\includegraphics[width=8cm]{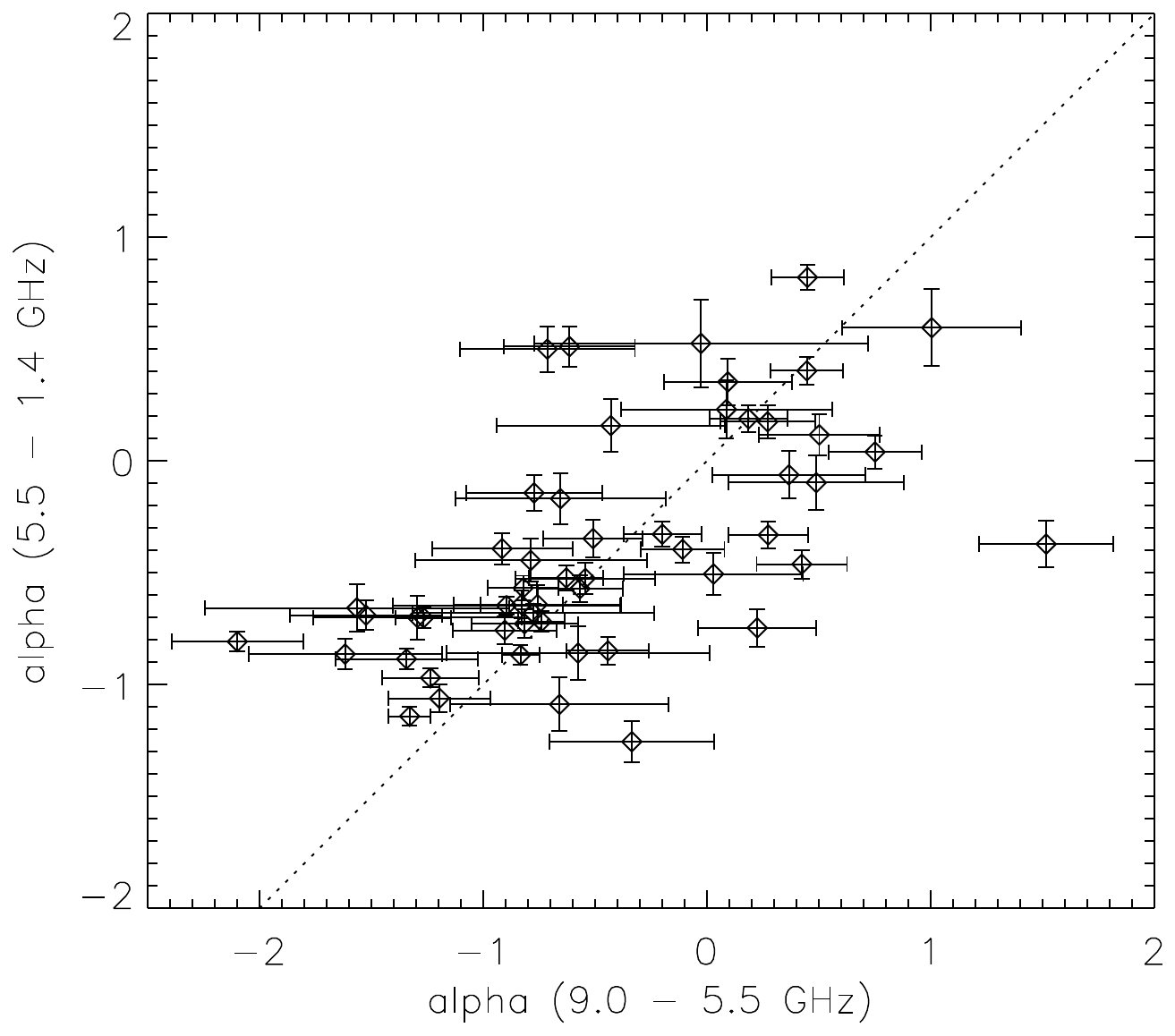}
\caption{The spectral index between 5.5 and 1.4 GHz versus spectral index between 9.0 and 5.5 GHz.} 
\label{im:alphavsalpha}
\end{figure}

The observed fraction of flat spectrum sources is similar to the results from H15 for the 5.5 GHz selected GHz sources in eCDFS. That work used the 1.4 to 5.5 GHz spectral index, whereas this result is from the 5.5 to 9.0 GHz spectral index, showing that in almost all cases the flat or inverted 5.5 GHz selected sources have continued to be flat or inverted at 9.0 GHz. This is confirmed in the plot of 1.4 to 5.5 GHz alpha vs 5.5 to 9.0 GHz alpha (Figure \ref{im:alphavsalpha}).  
Nearly all 9 GHz sources do not show a significant change from 1.4 to 9.0 GHz in their spectral index (Figure \ref{im:alphavsalpha}). We note however the uncertainty in the 5.5 to 9.0 GHz alpha can be large, due to the lower S/N of the sources in the 9.0 GHz images. Hence, more sensitive 9 GHz imaging would allow a better determination of whether there is significant change in the broadband radio SED in this frequency regime. Figure \ref{im:alphavsalpha} shows one outlier 9 GHz source, ID 59, which has a 1.4 to 5.5 GHz spectral index of $-0.37$ but inverted between 5.5 and 9 GHz with spectral index of 1.51 at higher frequencies. This source appears to have an ``upturned steep" radio SED \citep{harvey2018} which indicates that the high frequency observations may be detecting restarting radio jets while the lower frequency 1.4 GHz VLA observations are detecting the older lobes. This radio source is unresolved in the 1.4 GHz VLA imaging and 5.5 GHz ATCA imaging, and slightly extended in the 9 GHz imaging, so this may be a possibility. However, higher resolution imaging at 9 GHz combined with deeper 1.4 GHz imaging to pick up low surface brightness extended lobes would be needed to confirm this. 

Source ID 43 is a faint inverted spectrum source which was not detected in VLA imaging, down to 7.5 $\mu$Jy rms at 1.4 GHz, and is classified as a radio-quiet AGN. This implies that radio-quiet AGN may be more numerous than expected, even at these faint flux density levels. This inverted radio-quiet AGN is the radio-faint analogue to radio-loud high frequency peakers (HFPs), which are thought to be small and very young ($\lesssim$100 years old) radio sources \citep{dallacasa2003}. Next generation sky surveys such as Evolutionary Map of the Universe (EMU) on ASKAP \citep{norris2011}and VLASS are expected to reach 10 and 70 $\mu$Jy rms at 1.4 and 3.0 GHz, respectively, but faint inverted radio quiet AGN such as ID 43 will be completely missed by these surveys.  

Lastly, we note that time variability of radio sources can affect the measured spectral index but the 5.5 and 9.0 GHz ATCA data were obtained simultaneously so variability is not a factor in the measurements of the 5.5 to 9.0 GHz spectral indices.  

\section{Summary and Conclusions}

We have presented  Australia Telescope Compact Array 9.0 observations at 9.0 GHz of the extended Chandra Deep Field South. The resulting image of 0.276 square degrees reaches a sensitivity of $\sim$ 21 $\mu$Jy per beam rms, for a synthesised beam of 4.0 $\times$ 1.3 arcsec. This mosaic is the largest ever made at 9 GHz to these depths, and, importantly, the noise variation over the image is only about 20\%. We catalogued 55 individual radio sources and find 7 of these are multiple-component radio sources.

We calculated source counts at 9.0 GHz after careful corrections for completeness, flux boosting and resolution bias. These are amongst the deepest source counts in the 9 GHz band but come from an area much larger than previous VLA work, which consisted of one or two deep pointings. This work provides the best statistics for 0.1 $< S_{\rm 9 GHz} <$ 1 mJy. In general we find there is good agreement between the observed counts and the semi-empirical simulations of Wilman et al. 2008 (SKADS) and Bonaldi et al. 2019 (T-RECS). The T-RECS simulations predict about 50\% of the faintest sources in our sample would be star forming galaxies but we have detected only one star forming galaxy with our radio observations. Our observations do not have the surface brightness sensitivity to detect typical local star forming spirals, and they can account for this discrepancy. 

Using the wealth of multiwavelength data available in the eCDFS, we were able to classify the radio sources as AGN or star forming  and assign redshifts. Radio sources in eCDFS have been classified as radio-loud, radio-quiet or star forming by Bonzini et al. (2013), using MIR 24$\mu$m-radio flux density ratio, X-Ray luminosity and MIR (IRAC) colour-colour diagnostics. We applied this classification to 52/55 of the 9 GHz sources and examined the remaining three in the literature. We find 50/55 (91\%) of the 9 GHz sources are radio loud AGN, 4/55 (7\%) are radio quiet AGN, and only one source (2\%) is a star forming galaxy. Therefore, surveys even deeper than 100 $\mu$Jy, or with better surface brightness sensitivity, are required to probe the star forming population significantly at 9 GHz. Spectroscopic redshifts were available for 41 (75\%) of radio sources and 13 of the remaining 14 sources had a photometric redshift estimate. The observed redshift distribution shows a spike at $z \sim 0.7$ consistent with a well-known galaxy overdensity, and has an overall shape consistent with the T-RECS (Bonaldi et al. 2019) AGN model redshift distribution, but a larger sample size is required to draw definitive conclusions on the redshift distribution.  

Radio sources were matched to literature VLA 1.4 GHz data (Miller et al. 2013) and our ATCA observations at 5.5 GHz (H15). Nearly all 9 GHz sources do not show a significant change from 1.4 to 9.0 GHz in their spectral index (i.e. no significant curvature between 1.4 to 9.0 GHz). We find 20/55 or 36\% of the faint 9 GHz population are flat or inverted with $\alpha >  -0.3$. This fraction is greater than that seen in the AGN models of the T-RECS simulations suggesting the radio population models are missing faint (or low luminosity) flat and inverted spectrum sources at 9 GHz. One inverted spectrum source, ID 43, is a radio quiet AGN which is too faint to be detected in deep VLA imaging. This type of radio source, with estimated space density of $\sim$4 per square degree, would be missed in new all-sky surveys such as EMU and VLASS.

New deep and wide radio surveys at high frequencies ($\gtrsim$ 10 GHz) are expected over the next few years. ATCA is currently undertaking observations for a Legacy Project called the GAMA Legacy ATCA Southern Survey (GLASS, Huynh et al. in preparation). The GLASS legacy survey aims to cover the 50 square degrees of the GAMA G23 field \citep{driver2009,baldry2010} at 5.5 and 9.5 GHz to 30 and 50 $\mu$Jy rms, respectively, with 3000 hours of observations over 3 years. GLASS is expected to detect about 13,000 and 8,000 sources at 5.5 and 9.5 GHz, respectively, providing definitive source counts and a large sample for understanding the flat or inverted spectrum sources not detected at lower frequency (e.g. EMU and VLASS).  There is also an ATCA project to image the eCDFS at 8.5 GHz to $\sim$ 7 $\mu$Jy (project ID C3171, Galvin et al. in preparation). These deep observations are expected to detect about 10 starbursts to $z \sim 1$. The main goal of this project is to trace star formation to moderate redshift using 8.5 GHz, where the radio continuum starts to have a significant contribution from thermal radio emission. This may be a better tracer of instantaneous star formation rate than synchrotron emission at 1.4 GHz.  These new high frequency radio surveys will provide further valuable insights into radio AGN and star forming populations, their evolution, and constraints for future radio population modelling. 
 
\section*{Acknowledgements}

We thank the anonymous referee for their careful reading of the manuscript.  MH thanks A. Bonaldi for helpful discussions which have improved this paper. 

The Australia Telescope Compact Array is part of the Australia Telescope National Facility which is funded by the Australian Government for operation as a National Facility managed by CSIRO.
This research has made use of the VizieR catalogue access tool, CDS, Strasbourg, France (DOI : 10.26093/cds/vizier). This research has made use of the NASA/IPAC Extragalactic Database (NED), which is operated by the Jet Propulsion Laboratory, California Institute of Technology, under contract with the National Aeronautics and Space Administration.





\bibliographystyle{mnras}
\bibliography{refs}



\begin{table*}
\caption{The spectral indices between 1.4 and 9.0 GHz of the ATCA 9.0 GHz sample.}
\begin{threeparttable}
\begin{tabular}{lllllllcccc} \hline
ID & $S_{9.0 GHz}$  & $\delta S_{9.0 GHz}$  & $S_{5.5 GHz}$  & $\delta S_{5.5 GHz}$& $S_{1.4 GHz}$  & $\delta S_{1.4 GHz}$& $\alpha_{5.5 GHz}^{9.0 GHz}$ & $\delta$$\alpha_{5.5 GHz}^{9.0 GHz}$ & $\alpha_{1.4 GHz}^{5.5 GHz}$ & $\delta$$\alpha_{1.4 GHz}^{5.5 GHz}$  \\
 & ($\mu$Jy) & ($\mu$Jy) &  ($\mu$Jy) & ($\mu$Jy) &  ($\mu$Jy) & ($\mu$Jy) & & \\ \hline
     1 &    374 &     58 &    544 &     54 &      1310 &    19 & -0.76 &  0.37 & -0.64 &  0.07 \\
    2 &    386 &     52 &    505 &     37 &      1036 &    15 & -0.54 &  0.31 & -0.53 &  0.06 \\
    3 &    441 &     35 &    421 &     34 &       260 &    17 &  0.09 &  0.23 &  0.35 &  0.08 \\
    5 &    277 &     37 &    231 &     24 &       252 &    14 &  0.37 &  0.34 & -0.06 &  0.08 \\
    6 &    859 &     66 &    697 &     44 &      1312 &    13 &  0.42 &  0.20 & -0.46 &  0.05 \\
    7 &   1034 &     69 &   1286 &     80 &      4108 &    26 & -0.44 &  0.18 & -0.85 &  0.05 \\
    8 &    981 &     71 &   1298 &     80 &      2836 &    13 & -0.57 &  0.19 & -0.57 &  0.05 \\
    9 &    804 &     63 &    704 &     49 &       555 &    13 &  0.27 &  0.21 &  0.17 &  0.05 \\
   10 &   6749 &    364 &  10114 &    560 &     22000 &    13 & -0.82 &  0.16 & -0.57 &  0.04 \\
   11 &     95 &     24 &    127 &     13 &       410 &    15 & -0.58 &  0.56 & -0.86 &  0.08 \\
   12\tnote{a} &    235 &     36 &    499 &     33 &      1281 &    12 & -1.52 &  0.34 & -0.69 &  0.05 \\
   13 &    427 &     47 &    333 &     10 &       284 &    15 &  0.50 &  0.23 &  0.12 &  0.04 \\
   14 &   6842 &   1642 &  10638 &     56 &     25760 &    55 & -0.90 &  0.49 & -0.65 &  0.00 \\
   19 &   1086 &     20 &    949 &      9 &      1493 &    13 &  0.27 &  0.04 & -0.33 &  0.01 \\
   20 &    531 &     25 &    683 &     12 &      1098 &    51 & -0.51 &  0.10 & -0.35 &  0.04 \\
   21 &    212 &     42 &    203 &     24 &       148 &    12 &  0.09 &  0.47 &  0.23 &  0.10 \\
   22\tnote{a}  &    481 &     24 &    866 &     10 &      3700 &    29 & -1.20 &  0.10 & -1.06 &  0.01 \\
   23 &   1572 &    105 &   2366 &    146 &      5697 &    43 & -0.83 &  0.18 & -0.64 &  0.05 \\
   25\tnote{a}  &  13061 &   1045 &  24005 &     36 &     90450 &    60 & -1.24 &  0.16 & -0.97 &  0.00 \\
   29 &   7449 &   1788 &  10921 &     32 &     27710 &    67 & -0.78 &  0.49 & -0.68 &  0.00 \\
   33 &    169 &     23 &    103 &     15 &        46 &     8 &  1.00 &  0.40 &  0.59 &  0.17 \\
   34\tnote{a}  &    366 &     43 &   1702 &     86 &      4814 &   103 & -3.12 &  0.26 & -0.76 &  0.04 \\
   35 &   5269 &     32 &   7584 &     16 &     20240 &    62 & -0.74 &  0.01 & -0.72 &  0.00 \\
   36 &    119 &     27 &    147 &     17 &       118 &    13 & -0.43 &  0.51 &  0.16 &  0.12 \\
   37 &    118 &     18 &     92 &     11 &       106 &    13 &  0.49 &  0.39 & -0.10 &  0.12 \\
   38\tnote{a}  &   1254 &    136 &   3526 &     67 &     10650 &    48 & -2.10 &  0.22 & -0.81 &  0.01 \\
   40 &    288 &     35 &    422 &     36 &       514 &    12 & -0.77 &  0.30 & -0.14 &  0.06 \\
   41 &   1628 &    117 &   1718 &     99 &      2952 &    13 & -0.11 &  0.19 & -0.40 &  0.04 \\
   42 &    181 &     31 &    258 &     13 &       130 &    14 & -0.71 &  0.36 &  0.50 &  0.09 \\
   43 &    144 &     36 &     92 &     16 &       <30 &   --- &  0.91 &  0.62 &  >0.80 &   --- \\
   44 &     89 &     28 &     90 &     17 &        44 &     9 & -0.03 &  0.74 &  0.52 &  0.20 \\
   45\tnote{a}  &    191 &     34 &    424 &     13 &      1380 &    27 & -1.62 &  0.36 & -0.86 &  0.03 \\
   46 &  14823 &    842 &  11886 &    665 &      3871 &    13 &  0.45 &  0.16 &  0.82 &  0.04 \\
   47 &    185 &     32 &    183 &     18 &       365 &    13 &  0.03 &  0.40 & -0.51 &  0.08 \\
   48 &   1860 &    119 &   2052 &    116 &      3213 &    12 & -0.20 &  0.17 & -0.33 &  0.04 \\
   49 &    382 &     48 &    517 &     37 &       257 &    13 & -0.61 &  0.29 &  0.51 &  0.06 \\
   50 &   4400 &     23 &   3533 &    212 &      2037 &    13 &  0.45 &  0.12 &  0.40 &  0.04 \\
   51 &   5848 &    104 &  11253 &     23 &     53640 &    52 & -1.33 &  0.04 & -1.14 &  0.00 \\
   53 &   2648 &    166 &   2417 &    143 &      1868 &    14 &  0.19 &  0.17 &  0.19 &  0.04 \\
   54 &    962 &     74 &    665 &     46 &       630 &    15 &  0.75 &  0.21 &  0.04 &  0.05 \\
   55 &    355 &     49 &    558 &     40 &       954 &    14 & -0.92 &  0.31 & -0.39 &  0.05 \\
   56\tnote{a}  &    176 &     38 &    333 &     13 &       869 &    54 & -1.29 &  0.45 & -0.70 &  0.05 \\
   57 &    125 &     26 &    173 &     18 &       218 &    14 & -0.66 &  0.47 & -0.17 &  0.09 \\
   58 &    861 &     25 &   1173 &     12 &      2402 &    27 & -0.63 &  0.06 & -0.52 &  0.01 \\
   59\tnote{b}  &    392 &     24 &    186 &     19 &       310 &    14 &  1.51 &  0.25 & -0.37 &  0.08 \\
   60\tnote{a}  &    124 &     39 &    268 &     29 &       660 &    30 & -1.56 &  0.68 & -0.66 &  0.09 \\
   61\tnote{a}  &   2006 &    261 &   3886 &     25 &     13050 &    88 & -1.34 &  0.26 & -0.89 &  0.00 \\
   62 &    133 &     31 &    196 &     19 &       360 &    15 & -0.79 &  0.52 & -0.44 &  0.08 \\
   63 &    427 &     25 &    667 &     10 &      1880 &    17 & -0.90 &  0.12 & -0.76 &  0.01 \\
   64 &     97 &     19 &    134 &     20 &       596 &    16 & -0.66 &  0.49 & -1.09 &  0.11 \\
   65 &   8121 &     42 &  12243 &     23 &     40130 &    41 & -0.83 &  0.01 & -0.87 &  0.00 \\
   66 &    374 &     36 &    559 &      9 &      1513 &    14 & -0.82 &  0.20 & -0.73 &  0.01 \\
   67 &    416 &     68 &    491 &     15 &      2736 &   132 & -0.34 &  0.34 & -1.26 &  0.04 \\
   68\tnote{a}  &   2599 &     45 &   4849 &     62 &     12670 &    97 & -1.27 &  0.04 & -0.70 &  0.01 \\
   70 &    595 &     26 &    533 &     51 &      1479 &    18 &  0.22 &  0.21 & -0.75 &  0.07 \\ \hline
\end{tabular}
\label{tab:alphas}
\begin{tablenotes}
\item [a] spuriously steep source (some flux density is likely resolved out at 9 GHz or 9 GHz lobe(s) below detection limit)
\item [b] possible restarting source 
\end{tablenotes}
\end{threeparttable}
\end{table*}

\bsp	
\label{lastpage}
\end{document}